\documentclass[12pt,a4paper]{article}
\usepackage[english]{babel}
\usepackage[a4paper]{geometry}
\usepackage{lmodern,textcomp,amsmath,amssymb,mathtools,physics,comment,authblk,mathdots,cases,cite}
\usepackage{tikz,tikz-cd}
\usetikzlibrary{arrows.meta,automata,positioning,shadows,shapes.geometric}

\DeclareMathAlphabet{\mathcal}{U}{rsfs}{m}{n}

\usepackage[unicode=true,bookmarks=true,bookmarksopen=false,breaklinks=false,pdfborder={0 0 0},colorlinks=true]{hyperref}
\usepackage{xcolor}
\definecolor{cblue}{rgb}{0.16, 0.32, 0.75}
\definecolor{cred}{rgb}{0.7, 0.11, 0.11}
\hypersetup{%
	,linkcolor=cred
	,citecolor=cblue
	,urlcolor=black
}

\usepackage[normalem]{ulem}

\DeclareMathOperator{\diag}{diag}
\newcommand{\id}{\mathrm{I}} 
\newcommand{\iu}{\mathrm{i}\mkern1mu}

\newcommand{\J}{\mathrm{J}}
\newcommand{\gammac}{\,\check{\!\gamma}}
\newcommand{\gammat}{\,\tilde{\!\gamma}}
\newcommand{\gammab}{\,\bar{\!\gamma}}

\newcommand{\ols}[1]{\mskip.5\thinmuskip\overline{\mskip-.5\thinmuskip {#1} \mskip-.5\thinmuskip}\mskip.5\thinmuskip} 
\newcommand{\olsi}[1]{\,\overline{\!{#1}}} 
\makeatletter

\long\def\count@stringtoks#1{\tc@earg\count@toks{\string#1}}
\long\def\count@toks#1{\the\numexpr-1\count@@toks#1.\tc@endcnt}
\long\def\count@@toks#1#2\tc@endcnt{+1\tc@ifempty{#2}{\relax}{\count@@toks#2\tc@endcnt}}
\def\tc@ifempty#1{\tc@testxifx{\expandafter\relax\detokenize{#1}\relax}}
\long\def\tc@earg#1#2{\expandafter#1\expandafter{#2}}
\long\def\tctestifnum#1{\tctestifcon{\ifnum#1\relax}}
\long\def\tctestifcon#1{#1\expandafter\tc@exfirst\else\expandafter\tc@exsecond\fi}
\long\def\tc@testxifx{\tc@earg\tctestifx}
\long\def\tctestifx#1{\tctestifcon{\ifx#1}}
\long\def\tc@exfirst#1#2{#1}
\long\def\tc@exsecond#1#2{#2}
\makeatother

\title{\textbf{Dimensional reduction of the Dirac equation in arbitrary spatial dimensions}}

\author[$\hspace{0cm}$]{\normalsize Davide Lonigro$^{1,2}$}
\author[$\hspace{0cm}$]{Rocco Maggi$^{1,2}$}
\author[$\hspace{0cm}$]{Giuliano Angelone$^{1,2}$}
\author[$\hspace{0cm}$]{Elisa Ercolessi$^{3,4}$}
\author[$\hspace{0cm}$]{\\Paolo Facchi$^{1,2}$}
\author[$\hspace{0cm}$]{Giuseppe Marmo$^{5,6}$}
\author[$\hspace{0cm}$]{Saverio Pascazio$^{1,2}$}
\author[$\hspace{0cm}$]{Francesco V. Pepe$^{1,2}$}

\affil[$1$]{\small Dipartimento di Fisica and MECENAS, Universit\`a di Bari, I-70126 Bari, Italy}
\affil[$2$]{\small INFN, Sezione di Bari, I-70126 Bari, Italy}
\affil[$3$]{\small Dipartimento di Fisica e Astronomia, Universit\`a di Bologna, I-40127 Bologna, Italy}
\affil[$4$]{\small INFN, Sezione di Bologna, I-40127 Bologna, Italy}
\affil[$5$]{\small Dipartimento di Fisica, Universit\`a di Napoli Federico II, I-80126, Naples, Italy}
\affil[$6$]{\small INFN, Sezione di Napoli, I-80126, Naples, Italy}

\begin{document}
	
	\maketitle
	
	\begin{abstract}
		We investigate the general properties of the dimensional reduction of the Dirac theory, formulated in a Minkowski spacetime with an arbitrary number of spatial dimensions. This is done by applying Hadamard's method of descent, which consists in conceiving low-dimensional theories as a specialization of high-dimensional ones that are uniform along the additional space coordinate. We show that the Dirac equation reduces to either a single Dirac equation or two decoupled Dirac equations, depending on whether the higher-dimensional manifold has even or odd spatial dimensions, respectively. Furthermore, we construct and discuss an explicit hierarchy of representations in which this procedure becomes manifest and can easily be iterated.			
	\end{abstract}
	
	\section{Introduction}
	
	Besides its ``unreasonable effectiveness'' in the description of natural phenomena, as vividly depicted in Wigner's celebrated essay~\cite{wigner}, mathematical intuition can also pave the way to further advances in a physical theory, even to the point of accurately predicting the outcomes of experiments yet to be conceived. Quantum mechanics and quantum field theories brim with such predictions, a famous example being the quantum theory of the Lamb shift, whose first empirical observation was preceded by its mathematical formulation~\cite{dirac,Weinberg}.
	
	Following a similar line of thought, mathematicians and physicists often address questions that have long been the prerogative of philosophers and even novelists. An important instance is the problem of dimensionality. While the world accessible to our sensory experience has three spatial dimensions---namely, in a relativistic framework, it is described by a (3+1)-dimensional spacetime---many mathematical models that are employed in the natural sciences admit, sometimes in a rather ``natural'' way, generalizations to a lower or higher-dimensional spacetime. The question whether a $(3+1)$-dimensional spacetime can be somewhat justified on the ground of first principles, rather than on merely empirical grounds, is a closely related one; in this regard, Ehrenfest~\cite{ehrenfest} singled out some peculiarities of a $(3+1)$-dimensional space, among which it is worth mentioning the ``physical dualism'' between the electric and magnetic quantities (equal number of electric and magnetic field components), which, in his view, mirrored the ``geometrical dualism'' between rotations and translations (equal number of coordinate axes and coordinate planes).
	
	The Dirac equation represents a remarkable example of a theory of great physical interest which, at the same time, admits a natural generalization to a spacetime with an arbitrary number of dimensions, with the dimensionality entering the theory in a nontrivial and compelling way: namely, it determines the minimal number of internal degrees of freedom on the objects (\emph{spinors}) on which the Dirac differential operator should act.
	
	Dirac himself had noticed the possibility of casting his equation in a different spacetime. According to his own recollection~\cite{history}, in 1928 he was still pursuing a relativistic equation for the electron. He considered the Klein--Gordon equation unsatisfactory to this purpose, in particular since it lacked a probabilistic and a single-particle interpretation; in his view, such issues stemmed out of the fact that it implements an equality between \emph{squared} quantities, $E^2=\vb*p^2+m^2$, which determines a second-order equation. The correct equation should implement the equality $E=(\vb*p^2+m^2)^{1/2}$, leading to a first-order equation in the time derivative, such as the Schr\"odinger one. On the other hand, the relativistic nature of the Hamiltonian would require linearity in the momentum.
	
	Dirac's solution came in two steps. First (and less known), by ``playing about'' with the Pauli matrices, he stumbled upon the ``very pretty mathematical result''
	\begin{equation}\label{eq:pauli}
		\qty(\sigma_x p_x+\sigma_y p_y+\sigma_z p_z)^2=\vb*p^2\,,
	\end{equation}
	which had to ``be of importance'', because it ``provided effectively a method of taking the square root of the sum of three squares and getting it in a linear form'' \cite{history}. By this point, he had reduced the problem of extracting a linear square root of $\vb*p^2+m^2$, a sum of four squares, to finding out four objects, $\alpha_x,\alpha_y,\alpha_z,\beta$, which had to anticommute with each other and square to one, just like the Pauli matrices, yielding the similar equality
	\begin{equation}
		\qty(\alpha_x p_x+\alpha_y p_y+\alpha_z p_z+\beta m)^2=\vb*p^2+m^2\,.
	\end{equation}
	The second and widely renowned step is that, after struggling for a while with the Pauli matrices, Dirac saw no compelling reason for sticking to $2\times 2$ matrices, and realized that all the requirements could be attained with $4\times 4$ objects.
	
	One of the most striking remarks by Dirac is that, by ``replacing the $\sigma$-matrices by four-row-and-column matrices, one could easily take the square root of the sum of four squares, \emph{or even five squares if one wanted to} [emphasis added]''. Dirac essentially noticed that, from a mathematical standpoint, a fifth matrix of the same order may very well be added to the set, leading to a $(4+1)$-dimensional analogue of his new equation; in other words, the problems of writing a $(3+1)$- and a $(4+1)$-dimensional Dirac equation are tackled and solved together by his approach.
	
	The gamma matrices were introduced in the seminal 1928 paper~\cite{dirac28}, and described the fermionic character of the electron and its antiparticle. Dirac understood only later that the antiparticle would be a novel entity, yet to be discovered and to be named positron. Nowadays, Dirac's anticommuting matrices can be regarded in terms of irreducible representations of Clifford algebras; therefore, a natural notion of Dirac matrices and equation in arbitrary dimensions is available. In particular, a vast and detailed literature is devoted to the dimensional periodicity characterizing countless aspects of the problem: moving from very basic questions---such as the nature (complex or self-conjugate, real or quaternionic) of the representation, or the possibility of writing down purely real or purely imaginary matrices---an increasingly fine structure is revealed. When generalized to arbitrary dimensions, several familiar notions follow a pattern along some power of two: for instance, the chirality proceeds modulo~2, and the conjugations modulo~8~\cite{deWitSmith,Cornwell,Simon,pais,kawada}.
	
	In this work we address yet another facet of the problem: the link between the Dirac theories in different dimensions and their \emph{dimensional reduction} by descent. This technique, inspired by Hadamard's method of descent~\cite{hadamard,ehrenulen,balasz}, has been applied in~\cite{descent} as a simple method of conceiving the (2+1)- and (1+1)-dimensional electromagnetic theories as a specialization of the standard (3+1)-dimensional one to the case in which fields and sources do not depend on one or two spatial (Cartesian) coordinates. More recently, it has been applied to a Dirac particle, both free and minimally coupled to the electromagnetic field~\cite{descent2}; remarkably, in the interacting case the descent prescription can be ``upgraded'' to match a gauge-covariance request.
	
	Carrying forward this line of research, here we will discuss the dimensional reduction for the free Dirac equation in $d$ dimensions, with $d\geq2$ (hereafter, $d$ denotes the \emph{total} number of coordinates, including the temporal one). We will generalize and complete the results found in~\cite{descent2} on the Dirac equation in a (3+1)- and a (2+1)-dimensional spacetime, by showing that such a procedure is indeed compatible with the whole hierarchy of Dirac equations in all dimensions $d$, in the following sense:
	\begin{itemize}
		\item for odd $d$, the Dirac equation in $d$ dimensions reduces to a single Dirac equation in $(d-1)$ dimensions;
		\item for even $d$, the Dirac equation in $d$ dimensions decomposes into two decoupled Dirac equations in $(d-1)$ dimensions, pertaining to the two different classes of $(d-1)$-dimensional gamma matrices, which, for a suitable choice of representation, are mapped into each other via reflection of a single coordinate.
	\end{itemize}
	In this sense, when ``descending'' along the family of Dirac equations with different space dimensionality, a modulo 2 periodicity is observed. 
	
	This simple and elegant result can be obtained either via a more ``algebraic'' approach, or in a ``constructive'' way, that is, by constructing and working into suitable representations in which the picture underlined above is particularly simple. We regard these two perspectives as complementary. Moreover, the explicit representations we propose for handling the dimensional reduction procedure can be regarded as an adaptation of the Brauer--Weyl matrices~\cite{BrauerWeyl}, thus preserving the fundamental features of their construction.
	
	The paper is organized as follows. We start by recalling, in Section~\ref{sec:prelim}, the main properties of the Dirac theory for a free particle in a $d$-dimensional Minkowski spacetime, along with those of the Dirac matrices; in Section~\ref{sec:descent} we perform the dimensional reduction in the odd-to-even and even-to-odd cases with a representation-independent approach; in Section~\ref{sec:descent_explicit} we introduce suitable sets of Dirac matrices in any dimension in which the link between the equations under the descent is manifest; in Section~\ref{sec:DescRevisited} we review the descent in such representations. Final considerations and future research outlooks are reported in Section~\ref{sec:conclusions}.
	
	\section{Preliminaries}\label{sec:prelim}
	
	We shall consider $d$-dimensional Min\-kow\-skian spacetimes ($d\geq2$). Greek indices will refer to spacetime coordinates and vary between $0$ and $d-1$. The (global) coordinates will be denoted by $x=(x^\mu)$, $x^0$ being the only temporal one. The metric tensor will be $(\eta^{\mu\nu})=\diag(+1,-1,\dots,-1)$. When dealing at once with a $d$-dimensional spacetime and the lower-dimensional spacetimes obtained by dropping one or two spatial coordinates, their respective Min\-kow\-skian indices, free or contracted, will be primed according to the ``codimension'', that is, $0\le\mu\le d-1$, $0\le\mu'\le d-2$, $0\le\mu''\le d-3$. The identity matrix of order $n$ will be denoted by $\id_n$, or just $\id$ when there is no risk of confusion. Finally, the Kronecker product will be denoted by $\otimes$, and we will follow the convention that $A\otimes B$ is formally obtained by retaining the structure of the matrix $A$, and multiplying its elements by the matrix $B$.
	
	\subsection{Gamma matrices in arbitrary dimensions}\label{subsec:gamma}
	Let us start from $\mathrm{Cl}_{1,d-1}(\mathbb{R})$, the Min\-kow\-skian Clifford algebra of order $d$ ($d\ge2$)~\cite{Lee,Lounesto,SilvaRocha,park}, that is an associative unital algebra generated by the set $\{u^{\mu}\}_{\mu = 0,\dots,d-1}$ with the property 
	\begin{align}\label{eq:cliffordabstact}
		u^\mu u^\nu + u^\nu u^\mu =2\eta^{\mu\nu}\vb*1\,.
	\end{align}
	We regard a matrix representation of $\mathrm{Cl}_{1,d-1}(\mathbb{R})$	as an algebra of $N\times N$ complex matrices generated by the set $\{\gamma^\mu\}_{\mu = 0,\dots,d-1}$, such that Eq.~\eqref{eq:clifford} holds in the form
	\begin{align}\label{eq:clifford}
		\acomm{\gamma^\mu}{\gamma^\nu}=2\eta^{\mu\nu} \id_N\,,
	\end{align}
	that is, $\gamma^0$ squares to $\id_N$, each $\gamma^{j}$ ($j=1,\dots,d-1$) squares to $-\id_N$. Eq.~\eqref{eq:clifford} requires $N$ to be even, and each $\gamma^\mu$ to be traceless. Any family of $N\times N$ complex matrices $\{\gamma^\mu\}_{\mu = 0,\dots,d-1}$ which generates an irreducible representation (\textit{irrep}) of the Min\-kow\-skian Clifford algebra of order $d$ will be called a $d$-dimensional set of \emph{gamma} or \emph{Dirac matrices}. 
	
	Given a set of gamma matrices, and a strictly-increasing family of Min\-kow\-skian indices $0\le\mu_1<\dots<\mu_r\le d-1$, the ordered product $\gamma^{\mu_1}\dots\gamma^{\mu_r}$ will be denoted $\gamma^{\mu_1\dots\mu_r}$. The identity matrix can be regarded as the ordered product of $r=0$ matrices. From Eq.~\eqref{eq:clifford}, any product of gamma matrices can be written in the form $\pm\gamma^{\mu_1\dots\mu_r}$. In general, $\gamma^{\mu_1\dots\mu_r}$ is either involutive or anti-involutive, depending only on the number of its factors, and on whether $\gamma^0$ is one of them. For even $r$, $\gamma^{\mu_1\dots\mu_r}$ anticommutes with each of its factors and commutes with each $\gamma^\nu$ that is not a factor, and the converse is true for odd $r$.
	
	In even dimensions, the set $\{\gamma^{\mu_1\dots\mu_r}\}$ of all the ordered products is linearly independent. Since, for $d=2\ell$ ($\ell\ge1$), their number is $2^{2\ell}$, then $2^\ell$ is a lower bound on the order $N$ of the matrices. Actually, the $2\ell$-dimensional sets of gamma matrices are all and the only generating sets of irreps of order $2^\ell$, and the latter are all equivalent to each other (as algebra representations). In fact the ordered products of gamma matrices are a linear basis of the full matrix algebra $M_N(\mathbb{C})$, with $N=2^\ell$. If $\gamma^{\mu_1\dots\mu_r}\not\propto\id$, it anticommutes with at least one $\gamma^\mu$, hence is traceless. In particular, $\gamma^{0\dots 2\ell-1}$ anticommutes with each $\gamma^\mu$, and is traceless, and the same is true for the \emph{chiral matrix},
	\begin{equation}\label{eq:Chira}
		\gamma^{\mathrm{ch}}=\iu^{\ell-1} \, \gamma^{0\dots 2\ell-1}\,,
	\end{equation}
	which is also an involution,
	\begin{align}\label{eq:ChirInv}
		\qty(\gamma^{\mathrm{ch}})^2
		=\iu^{2(\ell-1)}\,
		\qty(-1)^{2\ell(2\ell-1)/2}
		\qty(\gamma^0)^2\dots \qty(\gamma^{2\ell-1})^2
		=\qty(-1)^{\qty(\ell-1)} \qty(-1)^{-\ell} (-1)^{2\ell-1}= \id\,.
	\end{align}
	
	In odd dimensions, the ordered products of the gamma matrices cannot be linearly independent: indeed, for $d=2\ell+1$ ($\ell\ge1$), $\gamma^{0\dots 2\ell}$ commutes with each $\gamma^\mu$, hence is proportional to the identity. Specifically, it can only be $\pm\iu^{\ell}\,\id$, since its square is
	\begin{align}
		\qty(\gamma^{0\dots 2\ell})^2
		=\qty(-1)^{2\ell(2\ell+1)/2}
		\,\qty(\gamma^0)^2\dots \qty(\gamma^{2\ell})^2
		=\qty(-1)^{\ell}\, \id\,.
	\end{align}
	This fact yields the irreducibility of the algebra generated by the subset $\{\gamma^{\mu'}\}$, obtained by dropping $\gamma^{2\ell}$, which is then a ($2\ell$)-dimensional set of gamma matrices. Consequently, the matrix order must be $N=2^\ell$, and the set of all the ordered products $\{\gamma^{\mu'_1\dots\mu'_r}\}$ not involving $\gamma^{2\ell}$ is a basis of the full matrix algebra $M_N(\mathbb{C})$. Moreover, by replacing $\gamma^{2\ell}$ with its opposite, a new ($2\ell+1$)-dimensional set $\{\gammac^\mu\}$ of  gamma matrices is obtained, differing from the initial one only by the sign of the last matrix. The sets $\{\gamma^\mu\}$ and $\{\gammac^\mu\}$ generate inequivalent representations (if it were $\gammac^{\mu}=S\gamma^\mu S^{-1}$, $S$ would commute with each $\gamma^{\mu'}$, so would be a scalar, and the two sets would coincide). Indeed, the ($2\ell+1$)-dimensional sets of gamma matrices are grouped precisely in two equivalence classes. All the ordered products different from $\id$ and $\gamma^{0\dots 2\ell}$ anticommute with at least one gamma matrix, hence are traceless.
	
	As a consequence of the discussion above, there are only two possible extensions of a $(2\ell)$-dimensional set $\{\gamma^{\mu'}\}$, with chiral matrix $\gamma^{\rm ch}$, to a $(2\ell+1)$-dimensional set, obtained by adding $\gamma^{2\ell}=\pm\iu\gamma^{\mathrm{ch}}$, the resulting sets being inequivalent	\cite{BrauerWeyl,Lee1,Lee,deWitSmith,Cornwell,Simon, Go55}.
	
	\subsection{Dirac theory in arbitrary dimensions}\label{subsec:dirac}
	Given a $d$-dimensional set $\{\gamma^\mu\}$ of gamma matrices, the Dirac equation is defined as follows~\cite{thaller}:
	\begin{equation}\label{eq:DiracEq}
		\left(\iu\gamma^\mu\partial_\mu-m\right)\Psi=0,
	\end{equation}
	where $m$ is the mass ($m\ge 0$), and $\Psi$, the \emph{spinor field}, is a vector-valued wa\-ve\-func\-tion $\Psi(x)\in\mathbb{C}^N$, with $N=2^{\lfloor d/2\rfloor}$, $x\in\mathbb{R}^{1+(d-1)}$ (hereafter, when $d$ is fixed, $N=2^{\lfloor d/2\rfloor}$ is always understood).
	It can be derived as the Euler--Lagrange equation of the \emph{Dirac Lagrangian}
	\begin{equation}\label{eq:DiracLagr}
		\mathcal{L}\qty(\Psi,\partial\Psi,\ols\Psi)
		=\ols{\Psi}\left(\iu\gamma^\mu\partial_\mu-m\right)\Psi,
	\end{equation}
	$\ols{\Psi}=\Psi^\dag\gamma^0$ being the \emph{Dirac adjoint} of the spinor $\Psi$. 
	
	The following condition
	\begin{equation}\label{eq:herm}
		\left(\gamma^\mu\right)^\dag=\gamma^0\gamma^\mu\gamma^0,
	\end{equation}
	ensures that the action defined by the Dirac Lagrangian is real. In the wake of the fundamental relations~\eqref{eq:clifford}, this condition is equivalent to requiring $\gamma^0$ to be Hermitian and each $\gamma^{j}$ to be anti-Hermitian,
	\begin{align}
		\left(\gamma^0\right)^\dag=\left(\gamma^0\right)^3=\gamma^0\,,
		&&
		\left(\gamma^j\right)^\dag=-\gamma^j\left(\gamma^0\right)^2=-\gamma^j\,;
	\end{align}
	or, equivalently, to requiring all the gamma matrices to be unitary,
	\begin{align}
		\gamma^0\left(\gamma^0\right)^\dag 
		=\left(\gamma^0\right)^4=\id\,,
		&&
		\gamma^j\left(\gamma^j\right)^\dag =-\left(\gamma^j\right)^2\left(\gamma^0\right)^2=\id\,.
	\end{align}
	Equivalent sets of unitary gamma matrices are related by unitary similarity transformations. In particular, modulo a phase, they are in a one-to-one correspondence with $\mathrm{U}(N)$. Herafter the condition~\eqref{eq:herm} will be assumed.
	
	The Dirac equation and the Dirac Lagrangian are manifestly invariant under transformations of the following kind:
	\begin{align}\label{eq:ReprChange}
		\gamma^{\mu}\to \gammat^{\mu}=U\gamma^{\mu}U^{\dagger}\,,
		&&\Psi\to \tilde\Psi=U\Psi\,,
	\end{align}
	with $U\in\mathrm{U}(N)$. We will say that that each $U$ defines a distinct \emph{representation} of the theory, and regard Eq.~\eqref{eq:ReprChange} as the general form of a representation change. Any ordered product $\gamma^{\mu_1\dots\mu_r}$ of gamma matrices trasforms like a gamma matrix:
	\begin{equation}\label{eq:ReprChangeProd}
		\gamma^{\mu_1\dots\mu_r}\to \gammat^{\mu_1\dots\mu_r}=U\gamma^{\mu_1\dots\mu_r}U^{\dagger}\,.
	\end{equation}
	
	This is true, in particular, for the chiral matrix in even dimensions. The chiral matrix is also Hermitian, because, if $d=2\ell$ ($\ell\ge1$),
	\begin{align}\label{eq:ChirHerm}
		\qty(\gamma^{\mathrm{ch}})^\dagger
		&= (-\iu)^{\ell-1} 
		(\gamma^0\gamma^{2\ell-1}\gamma^0)\dots (\gamma^0\gamma^1\gamma^0 )\gamma^0
		= (-1)^{\ell-1} (-1)^{(2\ell-1)(2\ell-2)/2}\gamma^{\mathrm{ch}}
		=\gamma^{\mathrm{ch}}\,.
	\end{align}
	Then it is unitary as well, being an involution.
	
	The $d$-dimensional spinor representation of the Lorentz group is generated by the set $\{\frac{\iu}{2}\,\gamma^{\mu\nu}\}$ of the ordered products of two gamma matrices,
	with $\frac{\iu}{2}\,\gamma^0\gamma^i$ generating the boosts along the $x^i$ direction ($1\le i\le d-1$), and $\frac{\iu}{2}\,\gamma^j\gamma^k$ the rotations in the $x^jx^k$ plane ($1\le j<k\le d-1$)~\cite{BrauerWeyl,deWitSmith}.
	
	\section{Representation-independent dimensional reduction}\label{sec:descent}
	
	In this section we will show that, by performing the descent on the Dirac equation in a $d$-dimensional spacetime ($d\ge 3$) along one spacelike direction, we obtain either one or two distinct Dirac equations for a $(d-1)$-dimensional spacetime, according to whether $d$ is even or odd. This procedure can be iterated until dimension 2 is reached. Quite interestingly, the reduction also works at the level of the Lagrangians.
	
	Without loss of generality, the descent from $d$ to $d-1$ dimensions will be performed by requiring the wa\-ve\-func\-tion to be independent of $x^{d-1}$, the ``last'' spatial coordinate,
	\begin{equation}\label{eq:DescentCondition}
		\partial_{d-1}\Psi=0\,.
	\end{equation}
	This condition is manifestly re\-pre\-sen\-ta\-tion-in\-de\-pen\-dent. It is also covariant with respect to the $(d-1)$-dimensional group of the $x^{d-1}$-preserving Lorentz transformations,
	\begin{align}\label{eq:lambda}
		\varLambda	
		=\left(\begin{array}{cc|c}
			\multicolumn{2}{c|}{L}&\vphantom{\biggl.\biggr.}\\
			\hline
			\phantom{1} & \phantom{1}& 1
		\end{array}\right)\,,
	\end{align}
	with $L\in\mathrm{O}(1,d-2)$. By imposing the descent condition~\eqref{eq:DescentCondition} on the Dirac equation~\eqref{eq:DiracEq}, we get the reduced equation
	\begin{equation}\label{eq:DiracRed}
		\qty(\iu \gamma^{\mu'}\partial_{\mu'}-m)\Psi=0\,,
	\end{equation}
	that is also the equation of motion of the reduced Lagrangian
	\begin{equation}\label{eq:redLagr}
		\mathcal{L}\qty(\Psi,\partial\Psi,\ols\Psi)\big|_{\partial_{d-1}\Psi=0}
		=\ols{\Psi}\left(\iu{\gamma^{\mu'}}\partial_{\mu'}-m\right)\Psi\,,
	\end{equation}
	obtained by restricting the Dirac Lagrangian~\eqref{eq:DiracLagr} to the wa\-ve\-func\-tions satisfying the descent condition~\eqref{eq:DescentCondition}.
	
	\subsection{Even-to-odd descent}\label{subsec:EvenOdd}
	This is the hard case, including as a notable instance the reduction of the ordinary Dirac equation from 4 to 3 spacetime dimensions that has been investigated in~\cite{descent2}. The treatment we will give here of the even-to-odd descent is a generalization of that special case.
	
	Let $d=2\ell$ ($\ell\ge1$) be the dimension of the spacetime of the initial theory. Recall that the gamma matrices in $2\ell$ and $2\ell-1$ dimensions have different orders, $N=2^{\ell}$ and $N/2=2^{\ell-1}$, respectively.
	
	\subsubsection{Decoupling}
	A projector on $\mathbb{C}^N$ is an idempotent linear map $P\colon\mathbb{C}^N\to \mathbb{C}^N$, $P=P^2$; $P$ is an orthogonal projector when it is also Hermitian, and a nontrivial projector when $P\ne0$, $P\ne\id_N$. In the latter case, $Q=\id_N-P$ is also a nontrivial projector, and is orthogonal if and only if $P$ is. By means of the resolution of the identity $P+Q=\id_N$, the reduced equation~\eqref{eq:DiracRed} can always be written as a coupled pair of projected equations
	\begin{numcases}{}
		P(\iu\gamma^{\mu'}\partial_{\mu'}-m)P\Psi
		+\iu P\gamma^{\mu'}Q \partial_{\mu'} \Psi=0\,,\\
		Q(\iu\gamma^{\mu'}\partial_{\mu'}-m)Q\Psi+\iu Q\gamma^{\mu'} P\partial_{\mu'}\Psi=0\,.
	\end{numcases}{}%
	Then the projected wavefunctions $P\Psi$ and $Q\Psi$ are decoupled when the coupling terms $\iu P\gamma^{\mu'} Q\partial_{\mu'}\Psi$ and $\iu Q\gamma^{\mu'} P\partial_{\mu'}\Psi$ vanish identically. A necessary and sufficient condition to this purpose is
	\begin{equation}\label{eq:projCommuto}
		\comm{P}{\gamma^{\mu'}}=0\,, \quad \text{for all }\mu'=0,\dots,2\ell-2.
	\end{equation}
	
	Let us look for a projector ensuring this condition. Since in even dimensions the set of the ordered products of the gamma matrices is a basis of the full matrix algebra, we can decompose $P$ (temporarily suspending Einstein summation convention) as
	\begin{equation}\label{eq:decomposition}
		P=a\id_N+\sum_\mu \,a_\mu\gamma^\mu+ 
		\sum_{\mu<\nu} \, a_{\mu\nu}\gamma^{\mu\nu}+
		\ldots+
		\;\sum_{\mathclap{\mu_1<\dots<\mu_p}} \, a_{\mu_1\dots\mu_p}\gamma^{\mu_1\dots\mu_p}+
		\ldots+a_{0\dots2\ell-1}\gamma^{0\dots2\ell-1}\,.
	\end{equation}
	Let us consider a generic term $a_{\mu_1\dots\mu_p}\gamma^{\mu_1\dots\mu_p}$ in the decomposition above.
	If $p=2r$, as long as $r\ne0$, at least two gamma matrices are involved, and $\mu_1\le2\ell-2$, so that $\comm{P}{\gamma^{\mu_1}}=0$, with $\gamma^{\mu_1}$ and $\gamma^{\mu_1\dots\mu_{2r}}$ anticommuting. On the other hand, if $p=2r+1$, we can find $\mu\notin\qty{\mu_1,\dots,\mu_{2r+1}}$, thus $\gamma^\mu$ anticommutes with $\gamma^{\mu_1\dots\mu_{2r+1}}$. Moreover, as long as $(\mu_1,\dots\mu_{2r+1})\ne(0,\dots,2\ell-2)$, we can choose $\mu\ne 2\ell-1$, so that $\comm{P}{\gamma^{\mu}}=0$. Therefore, for any term different from those along either $\id$ or $\gamma^{01\ldots 2\ell-2}$, we can find $0\le\nu\le2\ell-2$ such that $\comm{P}{\gamma^{\nu}}=0$ and $\acomm{\gamma^{\mu_1\dots\mu_p}}{\gamma^\nu}=0$. By computing the commutator by means of the decomposition~\eqref{eq:decomposition}, we get a vanishing linear combination of $\gamma^{\nu_1\dots\nu_q}\gamma^\nu$, which, modulo sign changes, is just a reshuffling of the basis vectors, with a contribution $\pm2 a_{\mu_1\dots\mu_p} \gamma^{\mu_1\dots\mu_p} \gamma^\nu$. In the end, only two terms of the decomposition~\eqref{eq:decomposition} can survive,
	\begin{equation}
		P=a\id_N+b\gamma^{2\ell-1}\gamma^{\rm ch}\,,
	\end{equation}
	with $\gamma^{\rm ch}$ the chiral matrix of the $(2\ell)$-dimensional set, as defined in Eq.~\eqref{eq:Chira}. This is the most generic operator ensuring the validity of Eq.~\eqref{eq:projCommuto}. Imposing $P^2=P$ yields a constraint on the coefficients,
	\begin{equation}
		a\id_N+b\gamma^{2\ell-1}\gamma^{\rm ch}
		=\qty(a^2+b^2)\id_N+2ab\gamma^{2\ell-1}\gamma^{\rm ch}\,,
	\end{equation}
	which singles out
	\begin{equation}\label{eq:KappaProj}
		P_\pm=\tfrac{1}{2}\qty(\id_N\mathrel{\pm}\kappa^{2\ell-1})\, ,
	\end{equation}
	with
	\begin{equation}\label{eq:kappa}
		\kappa^{2\ell-1}=\gamma^{2\ell-1}\gamma^{\mathrm{ch}}\,,
	\end{equation}	
	as the only nontrivial projectors satisfying Eq.~\eqref{eq:projCommuto}. The matrix $\kappa^{2\ell-1}$
	commutes with the first $2\ell-1$ gamma matrices and anticommutes with the last one
	\begin{align}\label{eq:KappaCommu}
		\comm{\gamma^{\mu'}}{\kappa^{2\ell-1}}=0\,,
		&&\acomm{\gamma^{2\ell-1}}{\kappa^{2\ell-1}}=0\,,
	\end{align}	
	is Hermitian (whence also involutive, unitary and traceless)
	\begin{align}\label{eq:KappaObs}
		\qty(\kappa^{2\ell-1})^\dagger
		=\gamma^{\mathrm{ch}}\gamma^0\gamma^{2\ell-1}\gamma^0
		=\gamma^{2\ell-1}\gamma^{\mathrm{ch}}
		=\kappa^{2\ell-1}\,,
	\end{align}
	and also transforms like the gamma matrices under a change of representation~\eqref{eq:ReprChange},
	\begin{equation}\label{eq:ReprChangeKappa}
		\kappa^{2\ell-1}\to {\,\tilde{\!\kappa}}^{2\ell-1}=U\kappa^{2\ell-1}U^{\dagger}\,,
	\end{equation}
	so that $\kappa^{2\ell-1}$ is Hermitian with eigenvalues $\pm1$, and the projectors~\eqref{eq:KappaProj} are precisely the orthogonal projectors onto the eigenspaces of $\kappa^{2\ell-1}$. Clearly they are mutually orthogonal, and resolve the identity. Moreover, they commute with all the $\gamma^{\mu'}$, and, in particular,
	\begin{align}
		P_\pm\gamma^{\mu'}P_\mp=0\,,&&P_\pm\gamma^{2\ell-1}P_\mp=0\,.
	\end{align}
	Thus, by setting
	\begin{align}\label{eq:Comp}
		\Psi_\pm=P_\pm\Psi\,,
		&&
		\gamma^{\mu'}_\pm=P_\pm\gamma^{\mu'} P_\pm\,,	
		&&
		\gamma^{2\ell-1}_{\pm,\mp}=P_\pm\gamma^{2\ell-1} P_\mp\,,	
	\end{align}
	the following decompositions hold:		
	\begin{align}\label{eq:DeComp}	
		\Psi=\Psi_+ + \Psi_-\,,
		&&
		\gamma^{\mu'}=\gamma^{\mu'}_+ + \gamma^{\mu'}_-\,,
		&&
		\gamma^{2\ell-1}=\gamma^{2\ell-1}_{+,-} + \gamma^{\mu'}_{-,+}\,,
	\end{align}
	with the sets of $N\times N$ matrices $\{\gamma_{\pm}^{\mu'}\}$ also satisfying the relations
	\begin{equation}\label{eq:CliffPlusMinus}
		\acomm{\gamma^{\mu'}_\pm}{\gamma^{\nu'}_\pm}
		=2\eta^{\mu'\nu'}P_\pm\,.
	\end{equation}	
	When the descent condition~\eqref{eq:DescentCondition} is applied,
	\begin{equation}\label{eq:DescentConditionEven}
		\partial_{2\ell-1}\Psi=0\,,
	\end{equation} 
	the reduced equation~\eqref{eq:DiracRed} splits into the independent projected equations
	\begin{align}
		\qty(\iu\gamma_{+}^{\mu'}\partial_{\mu'}-m)\Psi_{+}=0\,,
		\label{eq:DiracRedPlus}\\
		\qty(\iu\gamma_{-}^{\mu'}\partial_{\mu'}-m)\Psi_{-}=0\,.
		\label{eq:DiracRedMinus}
	\end{align}	
	
	By changing representations, we are free to diagonalize $\kappa^{2\ell-1}$ and its projectors as
	\begin{align}\label{eq:BlockDiagReprProj}
		\kappa^{2\ell-1}=\mqty(\id_{N/2}& 0\\0& -\id_{N/2} )\,,	
		&&
		P_{+}=\mqty(\id_{N/2}& 0\\0& 0 )\,,	
		&&
		P_{-}=\mqty(0& 0\\0&\id_{N/2})\,.
	\end{align}
	Then the decompositions~\eqref{eq:DeComp} consist in the block structures of the wa\-ve\-func\-tion,
	\begin{align}\label{eq:BlockDiagReprPsi}
		\Psi=\mqty(\psi_+\\ \psi_-)\,,
	\end{align}
	$\psi_\pm$ being column vectors with $N/2$ components, and of the Dirac matrices,
	\begin{align}\label{eq:BlockDiagReprGamma}
		\gamma^{\mu'}=\mqty(\Gamma^{\mu'}_+ & 0\\0&\Gamma^{\mu'}_-)\,,
		&&
		\gamma^{2\ell-1}=\mqty( 0&\Gamma^{2\ell-1}_{+,-}\\\Gamma^{2\ell-1}_{-,+}&0)\,,
	\end{align}
	the blocks $\Gamma^{\mu'}_\pm, \Gamma^{2\ell-1}_{\pm,\mp}$ being matrices of order $N/2$. The converse is also true: $\kappa^{2\ell-1}$ and its projectors satisfy Eq.~\eqref{eq:BlockDiagReprProj} whenever we choose a representation with gamma matrices in the form~\eqref{eq:BlockDiagReprGamma}. In particular, the existence of a nontrivial orthogonal projector commuting with all the $\gamma^{\mu'}$ is equivalent to their simultaneous block-dia\-go\-na\-li\-za\-bi\-li\-ty into two blocks of order $N/2$, and yields the block-off-dia\-go\-nal structure of $\gamma^{2\ell-1}$.
	
	As a consequence of the relations~\eqref{eq:CliffPlusMinus}, the sets $\{\Gamma^{\mu'}_\pm\}$ satisfy
	\begin{equation}\label{eq:CliffPlusMinusBis}
		\acomm{\Gamma_\pm^{\mu'}}{\Gamma_\pm^{\nu'}}
		=2\eta^{\mu'\nu'}\id_{N/2}\,.
	\end{equation}
	Since the gamma matrices in $2\ell-1$ dimensions are precisely of order $N/2$, $\{\Gamma^{\mu'}_\pm\}$ are both $(2\ell-1)$-dimensional sets of  gamma matrices, also inheriting the property~\eqref{eq:herm} from $\{\gamma^{\mu}\}$.
	
	Accordingly, the projected equations~\eqref{eq:DiracRedPlus}--\eqref{eq:DiracRedMinus} are the $(2\ell-1)$-dimensional Dirac equations
	\begin{align}
		\label{eq:GammaPlus}
		\qty(\iu\Gamma_{+}^{\mu'}\partial_{\mu'}-m)\psi_{+}=0\,,\\
		\qty(\iu\Gamma_{-}^{\mu'}\partial_{\mu'}-m)\psi_{-}=0\,.
		\label{eq:GammaMinus}
	\end{align}
	The even gamma sets in the form~\eqref{eq:BlockDiagReprGamma} provide a straightforward picture of the decoupling me\-cha\-nism brought about by the descent	(whence they will play a crucial role in Section~\ref{sec:descent_explicit}). 
	In fact, in any such representation, the Dirac equation~\eqref{eq:DiracEq} is written as
	\begin{align}\label{eq:EvenToOddDescent0}
		\left(\begin{array}{c|c}
			\iu\Gamma_+^{\mu'}\partial_{\mu'}-m
			&\iu\Gamma_{+,-}^{2\ell-1}\partial_{2\ell-1}\vphantom{\biggl.\biggr.}\\
			\hline
			\iu\Gamma_{-,+}^{2\ell-1}\partial_{2\ell}\vphantom{\biggl.\biggr.}
			&\iu\Gamma_-^{\mu'}\partial_{\mu'}-m
		\end{array}\right)
		\left(\begin{array}{c}\vphantom{\biggl.\biggr.}\psi_+\\
			\hline
			\vphantom{\biggl.\biggr.}\psi_-	\end{array}\right)=0\,.
	\end{align}		
	When the descent condition~\eqref{eq:DescentCondition} is imposed, the block-off-dia\-go\-nal terms of the Dirac operator are put to zero, and the decoupling~\eqref{eq:GammaPlus}--\eqref{eq:GammaMinus} is manifest:
	\begin{align}\label{eq:EvenToOddDescent}
		\left(
		\begin{array}{c|c}
			\iu\Gamma_+^{\mu'}\partial_{\mu'}-m
			&\vphantom{\biggl.\biggr.}\\
			\hline
			\vphantom{\biggl.\biggr.}
			&\iu\Gamma_-^{\mu'}\partial_{\mu'}-m
		\end{array}\right)
		\left(\begin{array}{c}\vphantom{\biggl.\biggr.}\psi_+\\
			\hline
			\vphantom{\biggl.\biggr.}\psi_-	\end{array}\right)=0\,.
	\end{align}
	The restricted covariance of the decompositions~\eqref{eq:DeComp}	as well as of the corresponding block structures~\eqref{eq:BlockDiagReprPsi}--\eqref{eq:BlockDiagReprGamma} is ensured  by the fact that $\kappa^{2\ell-1}$ is a scalar with respect to the $(2\ell-1)$-dimensional restricted Lorentz group~\eqref{eq:lambda}, that is, $\kappa^{2\ell-1}$ commutes with all its generators $\frac{\iu}{2}\,\gamma^{\mu'}\gamma^{\nu'}$. Therefore the $x^{2\ell-1}$-in\-de\-pen\-dent wa\-ve\-func\-tions $\psi_\pm$ depend on the same coordinates, and have the same number of components and Lorentz algebra as a $(2\ell-1)$-dimensional spinor field.
	
	The $\mathrm U (1)$ transformation $\Psi\to \mathrm{exp}(\iu\theta \kappa^{2\ell-1})\Psi$, while not a symmetry of the full Lagrangian~\eqref{eq:DiracLagr}, is a symmetry of the reduced Lagrangian~\eqref{eq:redLagr}. In even dimensions the descent condition~\eqref{eq:DescentCondition} forces the solution of the $2\ell$-dimensional reduced Dirac equation~\eqref{eq:DiracRed} into an eigenspace of $\kappa^{2\ell-1}$. This behavior is reminiscent of the massless case: the massless Lagrangian is symmetric with respect to the chiral transformations, and the massless solutions must be eigenstates of the chirality.
	
	The descent also works at the level of Lagrangians. Indeed, on the ground of the discussion above, the reduced Lagrangian is just the the sum of two independent Dirac Lagrangians in $2\ell-1$ dimensions,
	\begin{equation}\label{eq:DiracLagrDesc-even}
		\mathcal{L}\qty(\Psi,\partial\Psi,\ols\Psi)\big|_{\partial_{2\ell-1}\Psi=0}
		=\olsi{\psi}_+\left(\iu{\Gamma_+^{\mu'}}\partial_{\mu'}-m\right)\psi_+
		+\olsi{\psi}_-\left(\iu{\Gamma_-^{\mu'}}\partial_{\mu'}-m\right)\psi_-\,,
	\end{equation}
	the Euler--Lagrange equations of the low-dimensional Lagrangian being Eqs.~\eqref{eq:GammaPlus}--\eqref{eq:GammaMinus}.
	
	\subsubsection{Mathematical remarks}
	We point out an interesting feature of the block structure~\eqref{eq:BlockDiagReprGamma} of the $\gamma^{\mu'}$: the $(2\ell-1)$-dimensional sets $\{\Gamma^{\mu'}_+\}$ and $\{\Gamma^{\mu'}_-\}$ generate  inequivalent representations. To show this, let us consider the group $G$ generated by all the $\gamma^\mu$, made up of all the products of the gamma matrices with strictly increasing indices, and of their opposites, and its subgroup $G'$ generated by all the $\gamma^{\mu'}$, made up of the products not involving $\gamma^{2\ell-1}$. Each set  $\{\Gamma^{\mu'}_+\}$ and $\{\Gamma^{\mu'}_-\}$ is irreducible since its order is minimal, and induces a representation of $G'$ over $\mathbb{C}^N$, and these group representations are equivalent if and only if their respective algebra representations  are. Finally, by regarding $G$ as an irrep of itself over $\mathbb{C}^N$, we denote the characters of these three representations by $\chi_\pm$ and $\chi$. Since the sum of the square moduli of the characters of an irrep is equal to the group order, we have
	\begin{align}\label{eq:Char1}
		\sum_{g\in G} \abs{\chi(g)}^2=2^{2\ell+2}\,,&&\sum_{g\in G'} \abs{\chi_\pm(g)}^2=2^{2\ell+1}\,.
	\end{align}
	From the block structure~\eqref{eq:BlockDiagReprGamma}, the objects of $G'$ are direct sums of the representations induced by $\{\Gamma^{\mu'}_\pm\}$, those of $G\setminus G'$ are block-off-diagonal, hence traceless; therefore
	\begin{equation}\label{eq:Char2}
		\chi(g)=\begin{cases*}
			\chi_+(g)+\chi_-(g)\,,&for $g\in G'$,\\
			0\,, &for $g\in G\setminus G'$.
		\end{cases*}
	\end{equation}
	Putting Eqs.~\eqref{eq:Char1} and~\eqref{eq:Char2} together, we get the orthogonality relation
	\begin{equation}
		\sum_{g\in G'} \qty(\chi_+(g))^\ast \chi_-(g)=0\,,
	\end{equation}
	which proves the claim. The same result will be obtained far more easily in Section~\ref{sec:DescRevisited} within a re\-pre\-sen\-ta\-tion-de\-pen\-dent approach.
	As a consequence, the set  $\{\gamma^{\mu'}\}$ obtained by dropping the last matrix (which we know in advance to be a reducible representation of the Min\-kow\-skian Clifford algebra of order $2\ell-1$) decomposes into a direct sum of irreps with coefficients both equal to one.
	
	As to the block structure~\eqref{eq:BlockDiagReprGamma} of $\gamma^{2\ell-1}$, since the matrix is unitary and anti-involutive, its blocks are in turn unitary, and the opposite of the inverse of each other:
	\begin{align}\label{eq:blocksLast}
		\qty(\Gamma^{2\ell-1}_{\pm,\mp})^\dag=\qty(\Gamma^{2\ell-1}_{\pm,\mp})^{-1}
		=-\Gamma^{2\ell-1}_{\mp,\pm}\,.
	\end{align}
	Moreover, the off-block-diagonal nature of $\gamma^{2\ell-1}$ is inherited by the chiral matrix,
	\begin{align}\label{eq:BlockDiagReprChir}
		\gamma^{\rm ch}=\kappa^{2\ell-1}\gamma^{2\ell-1}
		=\mqty( 0&\Gamma^{2\ell-1}_{+,-}\\-\Gamma^{2\ell-1}_{-,+}&0)
		=\mqty( 0&\Gamma^{2\ell-1}_{+,-}\\\qty(\Gamma^{2\ell-1}_{+.-})^\dag&0)\,.
	\end{align}
	
	\subsection{Odd-to-even descent}\label{subsec:oddToEven}
	This is the simple case, a notable instance of which is the reduction from 3 to 2 dimensions.
	
	Let $d=2\ell+1$ ($\ell=1,2,\dots$) be the dimension of the spacetime of the initial theory. Now the   gamma matrices in $2\ell+1$ and $2\ell$ dimensions
	have the same order $N=2^{\ell}$. If $\{\gamma^\mu\}$ is a $(2\ell+1)$-dimensional set, its subset $\{\gamma^{\mu'}\}$, obtained by eliminating $\gamma^{2\ell}$, is a $(2\ell-1)$-dimensional set, also inheriting the property~\eqref{eq:herm}. 
	
	Therefore, when the descent condition~\eqref{eq:DescentCondition} is applied,
	\begin{equation}\label{eq:DescentConditionOdd}
		\partial_{2\ell}\Psi=0\,,
	\end{equation} 
	the $x^{2\ell}$-independent wa\-ve\-func\-tion $\Psi$ has the same number of components and the same Lorentz algebra of a $2\ell$-dimensional spinor field, while depending on the same coordinates and obeying the reduced equation~\eqref{eq:DiracRed}, which is a $2\ell$-dimensional Dirac equation.
	
	Again, the descent also ``works'' at the level of Lagrangians: in fact, the reduced Lagrangian~\eqref{eq:redLagr} is just an ordinary Dirac Lagrangian~\eqref{eq:DiracLagr} in $2\ell$ dimensions, with Euler--Lagrange equation the reduced equation~\eqref{eq:DiracRed}.
	
	\section{The adapted representations}\label{sec:descent_explicit}
	
	The reduction procedure of Section~\ref{sec:descent} can be exhaustively read in terms of gamma matrices in neighboring dimensions: given a $d$-dimensional set, the descent yields either one or two inequivalent $(d-1)$-dimensional  sets, according to whether $d$ is odd or even, such that the reduced equations of the former are precisely the ordinary equations of the latter. Thus, if we start from a $d$-dimensional set, we get the tree of its lower-dimensional descendants. 
	
	However, things are not that simple: while  the odd-to-even descent is straightforward, in general the even-to-odd descent involves a change of representation, in order to put the gamma matrices in the form of Eq.~\eqref{eq:BlockDiagReprGamma}. Therefore, when the descent is iterated, a redefinition of the matrices may be needed every two steps. It would be desirable to have, right from the start, a sequence of gamma sets, indexed according to the dimension $d\ge 2$, such that the $(d+1)$-dimensional set directly reduces to the $d$-dimensional one. If we were able to build such a sequence, then we could easily move along the whole hierarchy of the gamma sets.
	
	This goal will be achieved as follows. First, we observe that any even-dimensional set has a grandparent having the block structure~\eqref{eq:BlockDiagReprGamma}. Therefore, starting from any $2$-dimensional set with such a block structure, a sequence of even-dimensional ancestors with the same block structure can be recursively defined. Then, the voids corresponding to the odd dimensions can be filled, while also addressing the problem of the two odd-dimensional equivalence classes. Our procedure will lead to a slight modifications of the Brauer--Weyl (BW) matrices~\cite{BrauerWeyl}. Since we are essentially adapting the BW construction to the descent, they will be called \emph{adapted representations}.
	
	\subsection{Even dimensions: the iterative construction}\label{subsec:evenConstruction}
	In the following, in order to distinguish between gamma matrices with the same Min\-kow\-skian index but pertaining to different dimensions, a dimensional label will be attached to the matrices.
	
	\subsubsection{Induction step}	
	For $d=2\ell+2$ ($\ell\ge 1$), let $\{\gamma^{\mu''}_{(2\ell)}\}$ be any $2\ell$-dimensional set of  gamma matrices. We claim that the set $\{\gamma_{(2\ell+2)}^\mu\}$ defined by
	\begin{align}
		\label{eq:constr_even_Angelone1}
		\gamma^{\mu''}_{(2\ell+2)}
		&=\id_2\otimes\gamma^{\mu''}_{(2\ell)}
		=\left(\begin{array}{c|c}
			\gamma^{\mu''}_{(2\ell)}
			&\phantom{-\iu\gamma^{\mu''}_{(2\ell)}}\vphantom{\Bigl.\Bigr.}\\
			\hline
			\phantom{-\iu\gamma^{\mu''}_{(2\ell)}}\vphantom{\Bigl.\Bigr.}
			&\gamma^{\mu''}_{(2\ell)}
		\end{array}\right)\,,
		\\	
		\label{eq:constr_even_Angelone2}
		\gamma^{2\ell}_{(2\ell+2)}
		&=-\iu\sigma^3\otimes\gamma^{\rm ch}_{(2\ell)}
		=\left(\begin{array}{c|c}
			-\iu\gamma^{\rm ch}_{(2\ell)}
			&\phantom{-\iu\gamma^{\mu'}_{(2\ell)}}\vphantom{\Bigl.\Bigr.}\\
			\hline
			\phantom{-\iu\gamma^{\mu'}_{(2\ell)}}\vphantom{\Bigl.\Bigr.}
			&\iu\gamma^{\rm ch}_{(2\ell)}
		\end{array}\right)\,,
		\\	
		\label{eq:constr_even_Angelone3}
		\gamma^{2\ell+1}_{(2\ell+2)}
		&=\iu\sigma^2\otimes \gamma^{\rm ch}_{(2\ell)}
		=\left(\begin{array}{c|c}
			\phantom{-\iu\gamma^{\mu'}_{(2\ell)}}\vphantom{\Bigl.\Bigr.}
			&\gamma^{\rm ch}_{(2\ell)}\\
			\hline
			-\gamma^{\rm ch}_{(2\ell)}
			&\phantom{-\iu\gamma^{\mu'}_{(2\ell)}}\vphantom{\Bigl.\Bigr.}
		\end{array}\right)\,,			
	\end{align}
	is a $(2\ell+2)$-dimensional set of gamma matrices, with the block structure~\eqref{eq:BlockDiagReprGamma}, whose reduced (odd) sets $\{\gamma^{\mu'}_{(2\ell+1)}\}$ are reduced to the initial (even) set $\{\gamma^{\mu''}_{(2\ell)}\}$.
	
	To see that Eqs.~\eqref{eq:constr_even_Angelone1}--\eqref{eq:constr_even_Angelone3} define a set of gamma matrices, we first observe that the matrices are indeed of order $N=2^{\ell+1}$, and then, by temporarily renaming $\tau^1=-\sigma^3$ and $\tau^2=\sigma^2$,
	\begin{align}
		\acomm{\gamma_{(2\ell+2)}^{\mu}}{\gamma_{(2\ell+2)}^{\nu'}}
		&=\id_2\otimes\acomm{\gamma_{(2\ell)}^{\mu'}}{\gamma_{(2\ell)}^{\nu'}}
		=2\eta^{\mu'\nu'}\id_{2N}\,,\\
		\acomm{\gamma_{(2\ell+2)}^{\mu'}}{\gamma_{(2\ell+2)}^{2k-1+p}}
		&=\iu\tau^p\otimes
		\acomm{\gamma_{(2\ell)}^{\mu'}}{\gamma^{\rm ch}_{(2\ell)}}=0\,,\\
		\acomm{\gamma_{(2\ell+2)}^{2k-1+p}}{\gamma_{(2\ell+2)}^{2k-1+q}}
		&=\acomm{\iu\tau^p}{\iu\tau^q}\otimes\gamma^{\rm ch}_{(2\ell)}
		=-2\delta^{pq}\id_{2N}\,,
	\end{align}
	we obtain the fundamental relations~\eqref{eq:clifford}. By construction, the final set always has the block structure~\eqref{eq:BlockDiagReprGamma}, and property~\eqref{eq:herm} is hereditary. 
	
	Finally, $\{\gamma^{\mu}_{(2\ell+2)}\}$ is reduced to the sets
	\begin{align}\label{eq:OddPossibili}
		\left\{\gamma^{0}_{(2\ell)},\dots,\gamma^{2\ell-1}_{(2\ell)},-\iu\gamma^{\rm ch}_{(2\ell)}\right\}\,,
		&&
		\left\{\gamma^{0}_{(2\ell)},\dots,\gamma^{2\ell-1}_{(2\ell)},+\iu\gamma^{\rm ch}_{(2\ell)}\right\}\,,
	\end{align}
	differing only by the last matrix, which is eliminated by a further (odd-to-even) descent.
	
	\subsubsection{Base step}
	In order to obtain even-dimensional sets of gamma matrices satisfying the desired properties, we have to select a $2$-dimensional set of gamma matrices $\left\{\smash{\cramped{\gamma^0_{(2)}}},\smash{\cramped{\gamma^0_{(1)}}}\right\}$ being respectively diagonal and off-diagonal, as long as they anticommute and square to one. Such a choice is obviously not unique. Our choice is
	\begin{align}
		\label{eq:seed}
		\gamma_{(2)}^0=\sigma^3=\mqty( 1&0\\ 0 &-1)\,,
		&&\gamma_{(2)}^1=\iu\sigma^2=\mqty( 0&1\\ -1&0)\,.
	\end{align}
	The chiral matrix of the set is
	\begin{align}\label{eq:chir2}
		\gamma_{(2)}^{\mathrm{ch}}
		=\iu^{\frac{2}{2}-1}\gamma_{(2)}^0\gamma_{(2)}^1
		= \sigma^3 \qty(\iu\sigma^2)
		=\sigma^1
		=\mqty( 0&1\\ 1&0)\,.
	\end{align}
	We stress that the particular choice described above has been singled out so that the 4-dimensional set concided with the one used in~\cite{descent2} for the reduction of the ordinary Dirac equation.
	
	\subsubsection{Main properties}	
	A notable property of this construction is the sequence of the chiral matrices. In fact, the  chiral matrix in $2$ dimensions is $\sigma^1$, and chiral matrices of neighboring dimensions are related by
	\begin{align}\label{eq:chiraProg}
		\gamma^{\rm ch}_{(2\ell+2)}
		=\iu^{\ell}\gamma^0_{(2\ell+2)}\cdots\gamma^{2\ell+1}_{(2\ell+2)}
		=\iu^{\ell}
		(-\iu\sigma^3)(\iu\sigma^2)\otimes
		\gamma^0_{(2\ell)}\cdots\gamma^{2\ell-1}_{(2\ell)}\qty(\gamma^{\rm ch}_{(2\ell)})^2
		=\sigma^1\otimes\gamma^{\rm ch}_{(2\ell)}\,.
	\end{align}
	Therefore, the chiral matrix for $d=2\ell$ is precisely the $\ell$-th power of $\sigma^1$ with respect to the Kronecker product, 
	\begin{equation}\label{eq:chiralProgAngelone}
		\gamma^{\rm ch}_{(2\ell)}
		=\underbracket{\sigma^1\otimes\dots\otimes\sigma^1}_{\ell}\,=\J_N\,,
	\end{equation}
	where $\J_n$ denotes the \emph{exchange matrix} of order $n$,
	\begin{equation}
		\J_n=
		\left(\vphantom{\begin{matrix}1\\ \iddots\\ 1\end{matrix}}
		\right.\!
		\underbracket{\begin{matrix}
				&&1\\
				&\iddots&\\
				1&&
		\end{matrix}}_{n}
		\!\left.\vphantom{\begin{matrix}1\\ \iddots\\ 1\end{matrix}}\right)\,.
	\end{equation}
	Incidentally, we have just checked that the chiral matrix is in the form~\eqref{eq:BlockDiagReprChir}, and it is easy to see that the observable $\kappa^{2\ell-1}$, defined by Eq.~\eqref{eq:kappa}, is indeed in the form~\eqref{eq:BlockDiagReprProj}:	
	\begin{equation}\label{eq:kappa_explicit}
		\kappa^{2\ell-1}=\sigma^3\otimes\id_{N/2}
		=\begin{pmatrix}
			\id_{N/2}&\\&-\id_{N/2}
		\end{pmatrix}\,.
	\end{equation}
	The behavior of the adapted chiral matrices is reminiscent of the Brauer--Weyl (BW) matrices~\cite{BrauerWeyl}. This is not by chance: the Euclidean adapted even sequence, 
	\begin{align}
		\gammab_{(2\ell)}^{1}&=\gamma^{0}_{(2\ell)}\,,\\
		\gammab_{(2\ell)}^{j+1}&=\iu\gamma^j_{(2\ell)}\,,
	\end{align}
	is defined by the recurrence relation
	\begin{align}
		\gammab_{(2\ell+2)}^{j'}
		&=\id_2\otimes\gammab_{(2\ell)}^{j'}\,,\\
		\gammab_{(2\ell+2)}^{2\ell}
		&=\sigma^3\otimes\gammab_{(2\ell)}^{\rm ch}\,,\\
		\gammab_{(2\ell+2)}^{2\ell+1}
		&=-\sigma^2\otimes \gammab_{(2\ell)}^{\rm ch}\,,			
	\end{align}
	with initial step $(\sigma^3,-\sigma^2)$, bringing about 
	\begin{align}
		\begin{aligned}
			\gammab_{(2\ell)}^1
			&=\id_2\otimes\id_2\otimes\ldots\otimes\id_2\otimes\id_2\otimes\sigma^3\,,	\\	
			\gammab_{(2\ell)}^2
			&=\id_2\otimes\id_2\otimes\ldots\otimes\id_2\otimes\id_2\otimes(-\sigma^2)\,,\\
			\gammab_{(2\ell)}^3
			&=\id_2\otimes\id_2\otimes\ldots\otimes\id_2\otimes\sigma^3\otimes\sigma^1\,,	\\	
			\gammab_{(2\ell)}^4
			&=\id_2\otimes\id_2\otimes\ldots\otimes\id_2\otimes(-\sigma^2)\otimes\sigma^1\,,\\
			\mathmakebox[\widthof{$\gamma^4_{(2\ell)}$}]{\vdots}&\\
			\gammab_{(2\ell)}^{2\ell-1}
			&=\sigma^3\otimes\sigma^1\otimes\ldots\otimes\sigma^1\otimes\sigma^1\otimes\sigma^1\,,	\\	
			\gammab_{(2\ell)}^{2\ell}
			&=(-\sigma^2)\otimes\sigma^1\otimes\ldots\otimes\sigma^1\otimes\sigma^1\otimes\sigma^1\,.
		\end{aligned}
	\end{align}
	This is essentially a BW set with the Kronecker factors reshuffled by swapping the $s$-th factor with the  ($\ell-s$)-th one. In a sense, we are just taking a BW set, and reading it from the right to the left (or, which is the same, interpreting it by adopting the opposite convention on the Kronecker product). In this way, the adapted sets retain the fundamental features of the BW.
	
	\subsection{Odd dimensions: filling the voids}\label{subsec:oddConstruction}
	Each odd-dimensional adapted set is required to satisfy two conditions: (i) being a reduced set of the overlying adapted set, and (ii) reducing to the underlying one.
	
	For $d=2\ell+1$ ($\ell\ge1$), we know from the previous subsection that the first condition is satisfied only by the sets~\eqref{eq:OddPossibili}, and, from Section~\ref{sec:prelim}, that the only sets meeting the second condition are obtained by adding $\pm\iu\gamma^{\rm ch}_{(2\ell)}$ to the $2\ell$-dimensional adapted set. So the two requirements yield the same constraint: the $(2\ell+1)$-dimensional adapted set must be one of the sets~\eqref{eq:OddPossibili}. 
	We recall from Subsection~\ref{subsec:gamma} that the odd sets are grouped into two equivalence classes. Our candidates are inequivalent and differ by the sign of one matrix, and we have precisely the freedom to pick either class.
	
	In order to have a proper sequence (that is, one single set for each dimension), we finalize our construction by setting
	\begin{align}\label{eq:constr_odd_1}
		\gamma^{\mu'}_{(2\ell+1)}&=\gamma^{\mu'}_{(2\ell)}\,,\\
		\gamma^{2\ell}_{(2\ell+1)}&=-\iu\gamma^{\rm ch}_{(2\ell)}\,.
		\label{eq:constr_odd_2}
	\end{align}
	Obviously, the discarded option is recovered by changing the sign to the last matrix,
	\begin{align}\label{eq:oddInequi}
		\gammac^{\mu'}_{(2\ell+1)}=\gamma^{\mu'}_{(2\ell+1)}\,,&&
		\gammac^{2\ell}_{(2\ell+1)}=-\gamma^{2\ell}_{(2\ell+1)}\,.
	\end{align}
	The $(2\ell+1)$-dimensional  sets $\{\gamma^{\mu'}_{(2\ell+1)}\}$ and $\{\gammac^{\mu'}_{(2\ell+1)}\}$ are inequivalent, and complementary within the $(2\ell+2)$-dimensional adapted set, in the sense of Eqs.~\eqref{eq:constr_even_Angelone1}--\eqref{eq:constr_even_Angelone2},
	\begin{align}\label{eq:complement}
		\gamma_{(2\ell+2)}^{\mu'}
		=\left(\begin{array}{c|c}
			\gamma_{(2\ell+1)}^{\mu'}
			&\vphantom{\Bigl.\Bigr.}\\
			\hline
			\vphantom{\Bigl.\Bigr.}
			&	\gammac_{(2\ell+1)}^{\mu'}
		\end{array}\right)\,,
		&&
		\gamma_{(2\ell+2)}^{2\ell-1}
		=\left(\begin{array}{c|c}
			\vphantom{\Bigl.\Bigr.}
			&\iu\gamma_{(2\ell+1)}^{2\ell}\\
			\hline
			\iu\gammac_{(2\ell+1)}^{2\ell}
			&\vphantom{\Bigl.\Bigr.}
		\end{array}\right)
		\,.
	\end{align}
	
	\subsection{Visual construction}\label{subsec:Visualia}				
	Here we shall provide a graphical interpretation of the first steps of our construction. From now on, the only numbers we are going to deal with will be 0 and the fourth roots of unity, $\pm1$, $\pm\iu$, which together make up a multiplicative monoid. Accordingly, we set
	\begin{align}
		\raisebox{-1.5pt}{\tikz{	\begin{scope}[scale=.36]
	\draw[fill=white,draw=black,very thin] (0,0) rectangle (1,1);
\end{scope}		}}=0\,,			
		&&			
		\raisebox{-1.5pt}{\tikz{	\begin{scope}[scale=.36]
	\draw[fill=blue!50!black!,
	draw=black,very thin] (0,0) rectangle (1,1);
\end{scope}		}}=1\,,			
		&&
		\raisebox{-1.5pt}{\tikz{	\begin{scope}[scale=.36]
						\draw[fill=red!50!black!,
						draw=black,very thin] (0,0) rectangle (1,1);
	\end{scope}		}}=-1\,,			
		&&			
		\raisebox{-1.5pt}{\tikz{	\begin{scope}[scale=.36]
	\draw[fill=green!50!black!,
	draw=black,very thin] (0,0) rectangle (1,1);
\end{scope}		}}=\iu\,,			
		&&
		\raisebox{-1.5pt}{\tikz{	\begin{scope}[scale=.36]
	\draw[fill=orange!75!black!,
	draw=black,very thin] (0,0) rectangle (1,1);
\end{scope}		}}=-\iu\,.			
	\end{align}
	and translate the multiplication table into the following composition law of colors:
	\begin{equation}
		\begin{aligned}
			\raisebox{-1.5pt}{\tikz{	\begin{scope}[scale=.36]
		
\begin{scope}[shift={(1,0)}]						
	\fill[blue!50!black] (0,5) rectangle (1,4);
	\fill[red!50!black] (1,5) rectangle (2,4);
	\fill[green!50!black] (2,5) rectangle (3,4);
	\fill[orange!75!black] (3,5) rectangle (4,4);
\end{scope}				

\begin{scope}[shift={(0,-1)}]			
	
	\fill[blue!50!black] (-1,4) rectangle (0,3);
	\fill[red!50!black] (-1,3) rectangle (0,2);
	\fill[green!50!black] (-1,2) rectangle (0,1);
	\fill[orange!75!black] (-1,1) rectangle (0,0);
	
\end{scope}				

\begin{scope}[shift={(1,-1)}]				
	\begin{scope}[shift={(0,-1)}]
		\fill[blue!50!black] (0,5) rectangle (1,4);
		\fill[red!50!black] (1,5) rectangle (2,4);
		\fill[green!50!black] (2,5) rectangle (3,4);
		\fill[orange!75!black] (3,5) rectangle (4,4);
		
	\end{scope}		
	
	\begin{scope}[shift={(0,-2)}]
		\fill[red!50!black] (0,5) rectangle (1,4);
		\fill[blue!50!black] (1,5) rectangle (2,4);
		\fill[orange!75!black] (2,5) rectangle (3,4);
		\fill[green!50!black] (3,5) rectangle (4,4);
		
	\end{scope}		
	
	\begin{scope}[shift={(0,-3)}]
		\fill[green!50!black] (0,5) rectangle (1,4);
		\fill[orange!75!black] (1,5) rectangle (2,4);
		\fill[red!50!black] (2,5) rectangle (3,4);
		\fill[blue!50!black] (3,5) rectangle (4,4);
		
	\end{scope}		
	
	\begin{scope}[shift={(0,-4)}]
		\fill[orange!75!black] (0,5) rectangle (1,4);
		\fill[green!50!black] (1,5) rectangle (2,4);
		\fill[blue!50!black] (2,5) rectangle (3,4);
		\fill[red!50!black] (3,5) rectangle (4,4);
		
	\end{scope}

\end{scope}					

\draw[step=1cm,teal!70!black!40,thin] (-1,3) -- (5,3);
\draw[step=1cm,teal!70!black!40,thin] (-1,2) -- (5,2);
\draw[step=1cm,teal!70!black!40,thin] (-1,1) -- (5,1);
\draw[step=1cm,teal!70!black!40,thin] (-1,0) -- (5,0);
\draw[thin,teal!70!black!40] (0,-1) -- (0,4) [rounded corners=.4pt]-- (-1,4)-- (-1,-1) [sharp corners]--cycle;

\draw[step=1cm,teal!70!black!40,thin] (0,-1) -- (0,5);
\draw[step=1cm,teal!70!black!40,thin] (1,-1) -- (1,5);
\draw[step=1cm,teal!70!black!40,thin] (2,-1) -- (2,5);
\draw[step=1cm,teal!70!black!40,thin] (3,-1) -- (3,5);
\draw[step=1cm,teal!70!black!40,thin] (4,-1) -- (4,5);
\draw[thin,teal!70!black!40] (0,4) -- (5,4) [rounded corners=.4pt]-- (5,5)-- (0,5) [sharp corners]--cycle;

\draw[thin,teal!70!black!40] (0,-1) --(0,4) -- (5,4) [rounded corners=.4pt]-- (5,-1) [sharp corners]-- cycle;

\draw[ultra thick,teal!70!black!40,rounded corners=1.9pt,line cap=round] (-1,4) --(0,4) -- (0,5);
\draw[ultra thick,teal!70!black!40,rounded corners=1.9pt,line cap=round] (0,-1) -- (0,4)--(5,4);

\end{scope}		}}
		\end{aligned}\;.
	\end{equation}
	Since the chiral matrix of an even adapted set is known in advance (see Eq.~\eqref{eq:chiralProgAngelone}), there will be no need to perform any matrix product. From now on, only Kronecker products will be involved, and this fact ultimately allows us to construct our gamma matrices by just ``playing with colors and shapes''.
	
	Our construction is entirely determined by 6 matrices of order 2, namely: the  gamma matrices in 2 dimensions, their chiral matrix, as well as the left factors of the Kronecker products~\eqref{eq:constr_even_Angelone1}--\eqref{eq:constr_even_Angelone3}. Each adapted gamma matrix is a Kronecker product of such fundamental objects. The left factors in Eqs.~\eqref{eq:constr_even_Angelone1}--\eqref{eq:constr_even_Angelone3} are $\id_2$, $-\iu\sigma^3$, and $\iu\sigma^2$, that is,
	\begin{align}\label{eq:left}
		\begin{aligned}
			\raisebox{-1.5pt}{\tikz{\begin{scope}[scale=.36]
	
				\draw [
		ultra thick,cyan!70!black!50,rounded corners=1pt]  (-.3,2.3) -- (-.3,-.3) --(8.3,-.3) --(8.3,2.3) --cycle;

	\fill[white] (0,2) rectangle (2,0);
	\fill[blue!50!black] (1,0) rectangle (2,1);
	\fill[blue!50!black] (0,1) rectangle (1,2);			
	
	\draw[step=1cm,black,very thin] (0,0) grid (2,2);	
	
	\begin{scope}[shift={(3,0)}]
		
		\fill[white] (0,2) rectangle (2,0);
		\fill[green!50!black] (1,0) rectangle (2,1);
		\fill[orange!75!black] (0,1) rectangle (1,2);			
		
		\draw[step=1cm,black,very thin] (0,0) grid (2,2);
	\end{scope}

	\begin{scope}[shift={(6,0)}]
		
		\fill[red!50!black] (0,0) rectangle (1,1);
		\fill[blue!50!black] (1,1) rectangle (2,2);
		
		\draw[step=1cm,black,very thin] (0,0) grid (2,2);
	\end{scope}		
\end{scope}}}
		\end{aligned}\,.
	\end{align}
	
	\paragraph{}
	\noindent{\bf The 2-dimensional set}, complete of its chiral matrix, is given by Eqs.~\eqref{eq:seed}--\eqref{eq:chir2},
	\begin{align}\label{eq:2dimVerbo}
		\gamma_{(2)}^0=\sigma^3\,,
		&&
		\gamma_{(2)}^1=\iu\sigma^2\,,
		&&
		\gamma_{(2)}^{\mathrm{ch}}=\sigma^1\,,
	\end{align}
	that is,
	\begin{align}\label{eq:2dim}
		\begin{aligned}
			\raisebox{-1.5pt}{\tikz{\begin{scope}[scale=.36]

	
	\draw [
	ultra thick,purple!70!black!40,rounded corners=1pt]  (-.3,2.3) -- (-.3,-.3) --(5.3,-.3) --(5.3,2.3) --cycle;


	\fill[white] (0,2) rectangle (2,0);
	\fill[red!50!black] (1,0) rectangle (2,1);
	\fill[blue!50!black] (0,1) rectangle (1,2);			
	
	\draw[step=1cm,black,very thin] (0,0) grid (2,2);	
	
	\begin{scope}[shift={(3,0)}]
		
		\fill[white] (0,2) rectangle (2,0);
		\fill[red!50!black] (0,0) rectangle (1,1);
		\fill[blue!50!black] (1,1) rectangle (2,2);
		
		\draw[step=1cm,black,very thin] (0,0) grid (2,2);
	\end{scope}

	\begin{scope}[shift={(6,0)}]
		
		\fill[blue!50!black] (0,0) rectangle (1,1);
		\fill[blue!50!black] (1,1) rectangle (2,2);
		
		\draw[step=1cm,black,very thin] (0,0) grid (2,2);
	\end{scope}		
\end{scope}}}
		\end{aligned}\,,
	\end{align}
	where we put the gamma matrices in the box and the chiral matrix outside.
	\paragraph{}
	\noindent{\bf The 3-dimensional set} is just the 2-dimensional one plus $-\iu$ times its chiral matrix,
	\begin{align}\label{eq:3dimVerbo}
		\gamma_{(3)}^0=\sigma^3\,,
		&&
		\gamma_{(3)}^1=\iu\sigma^2\,,
		&&
		\gamma_{(3)}^2=-\iu\sigma^1\,,
	\end{align}
	that is, we take the set~\eqref{eq:2dim} and just multiply the matrix outside the contour by $-\iu$,
	\begin{align}\label{eq:3dim}
		\begin{aligned}
			\raisebox{-1.5pt}{\tikz{	\begin{scope}[scale=.36]

	\fill[white] (0,2) rectangle (2,0);
	\fill[red!50!black] (1,0) rectangle (2,1);
	\fill[blue!50!black] (0,1) rectangle (1,2);			
	
	\draw[step=1cm,black,very thin] (0,0) grid (2,2);	
	
	\begin{scope}[shift={(3,0)}]
		
		\fill[white] (0,2) rectangle (2,0);
		\fill[red!50!black] (0,0) rectangle (1,1);
		\fill[blue!50!black] (1,1) rectangle (2,2);
		
		\draw[step=1cm,black,very thin] (0,0) grid (2,2);
	\end{scope}

	\begin{scope}[shift={(6,0)}]
		
		\fill[orange!75!black] (0,0) rectangle (1,1);
		\fill[orange!75!black] (1,1) rectangle (2,2);
		
		\draw[step=1cm,black,very thin] (0,0) grid (2,2);
	\end{scope}

\end{scope}}}
		\end{aligned}\,.
	\end{align}
	
	\paragraph{}
	\noindent{\bf The 4-dimensional set} is made up of four Kronecker products: the first two matrices come from the products between $\id_2$ and the matrices in 2 dimensions, the last two from the products between the fundamental matrices $-\iu\sigma^3$ and $\iu\sigma^2$ and $\sigma^1$, the  chiral matrix in 2 dimensions. Therefore, we consider the $3\times3$ multiplicative table of the Kronecker product, with column entries~\eqref{eq:left}, and row entries~\eqref{eq:2dim}. We only need the products in the first row and in the last column, without the one placed at their intersection:
	\begin{align}
		\begin{aligned}
			\raisebox{-1.5pt}{\tikz{	\begin{scope}[scale=.36]

	
	\draw[very thin,fill=gray!20!white] (-0.5,-0.5) -- (14.5,-0.5) [rounded corners]-- (14.5,2.5) -- (-0.5,2.5) [sharp corners]-- cycle;
	\draw[very thin,fill=gray!20!white] (-0.5,-0.5) -- (-0.5,-15.5) [rounded corners]-- (-5.5,-15.5) -- (-5.5,-.5) [sharp corners]-- cycle;

	\draw[help lines,step=5,xshift=-.5cm,yshift=-.5cm] (0,0) grid (15,-15);


	\begin{scope}[shift={(1,0)}]
		
		\draw [
		ultra thick,purple!70!black!40,rounded corners=1pt] (-.3,-.3) -- (-.3,2.3) --(7.3,2.3) --(7.3,-.3) --cycle;

		\fill[white] (0,2) rectangle (2,0);
		\fill[red!50!black] (1,0) rectangle (2,1);
		\fill[blue!50!black] (0,1) rectangle (1,2);			
		
		\draw[step=1cm,black,very thin] (0,0) grid (2,2);	
		
		\begin{scope}[shift={(5,0)}]
			
			\fill[white] (0,2) rectangle (2,0);
			\fill[red!50!black] (0,0) rectangle (1,1);
			\fill[blue!50!black] (1,1) rectangle (2,2);
			
			\draw[step=1cm,black,very thin] (0,0) grid (2,2);
		\end{scope}

		\begin{scope}[shift={(10,0)}]
			\fill[white] (0,2) rectangle (2,0);
			\fill[blue!50!black] (0,0) rectangle (1,1);
			\fill[blue!50!black] (1,1) rectangle (2,2);
			
			\draw[step=1cm,black,very thin] (0,0) grid (2,2);
		\end{scope}		
	\end{scope}

	
	\begin{scope}[shift={(-8,-2)}]
		\draw[ultra thick,cyan!70!black!50,rounded corners=1pt]  (-.3,.3) -- (2.3,.3) --(2.3,-12.3) --(-.3,-12.3) --cycle;
	\end{scope}	
	
	

	\begin{scope}[shift={(-5,-5)}]
		
		\fill[white] (0,0) rectangle (4,4);
		
		\draw[step=2cm,black,very thin] (0,0) grid (4,4);
		
		\draw [fill=blue!50!black!,
		draw=blue!50!black!, thick,rounded corners=1pt,fill opacity=0.15](0,2) rectangle (2,4);
		
		\draw [fill=blue!50!black!,
		draw=blue!50!black!, thick,rounded corners=1pt,fill opacity=0.15](2,0) rectangle (4,2);	
		
	\end{scope}

	\begin{scope}[shift={(-5,-10)}]

		\fill[white] (0,0) rectangle (4,4);
		
		\draw[step=2cm,black,very thin] (0,0) grid (4,4);
		
		\draw [fill=orange!75!black!,
		draw=orange!75!black!, thick,rounded corners=1pt,fill opacity=0.15](0,2) rectangle (2,4);
		
		\draw [fill=green!50!black!,
		draw=green!50!black!, thick,rounded corners=1pt,fill opacity=0.15](2,0) rectangle (4,2);	
		
	\end{scope}

	\begin{scope}[shift={(-5,-15)}]
		
		\fill[white] (0,0) rectangle (4,4);
		
		\draw[step=2cm,black,very thin] (0,0) grid (4,4);
		
		\draw [fill=blue!50!black!,
		draw=blue!50!black!, thick,rounded corners=1pt,fill opacity=0.15](2,2) rectangle (4,4);
		
		\draw [fill=red!50!black!,
		draw=red!50!black!, thick,rounded corners=1pt,fill opacity=0.15](0,0) rectangle (2,2);	
		
	\end{scope}

	\begin{scope}[shift={(-8,-4)}]
		
		\fill[blue!50!black] (1,0) rectangle (2,1);
		\fill[blue!50!black] (0,1) rectangle (1,2);
		
		\draw[step=1cm,black,very thin] (0,0) grid (2,2);
		
		\draw[ultra thin, gray] (0,2) -- (3,3);	
		\draw[ultra thin, gray] (2,2) -- (7,3);				
		\draw[ultra thin, gray] (2,0) -- (7,-1);				
		\draw[ultra thin, gray] (0,0) -- (3,-1);

	\end{scope}

	\begin{scope}[shift={(-8,-9)}]
		
		\fill[green!50!black] (1,0) rectangle (2,1);
		\fill[orange!75!black] (0,1) rectangle (1,2);
		
		\draw[step=1cm,black,very thin] (0,0) grid (2,2);
		
		\draw[ultra thin, gray] (0,2) -- (3,3);	
		\draw[ultra thin, gray] (2,2) -- (7,3);				
		\draw[ultra thin, gray] (2,0) -- (7,-1);				
		\draw[ultra thin, gray] (0,0) -- (3,-1);

	\end{scope}		
	
	\begin{scope}[shift={(-8,-14)}]
		
		\fill[red!50!black] (0,0) rectangle (1,1);
		\fill[blue!50!black] (1,1) rectangle (2,2);
		
		\draw[step=1cm,black,very thin] (0,0) grid (2,2);
		
		\draw[ultra thin, gray] (0,2) -- (3,3);	
		\draw[ultra thin, gray] (2,2) -- (7,3);				
		\draw[ultra thin, gray] (2,0) -- (7,-1);				
		\draw[ultra thin, gray] (0,0) -- (3,-1);		
		
	\end{scope}

	
	\begin{scope}[shift={(0,-5)}]
		
		\begin{scope}[shift={(0,2)}]
			\fill[red!50!black] (1,0) rectangle (2,1);
			\fill[blue!50!black] (0,1) rectangle (1,2);
			
			\draw[step=1cm,black,very thin] (0,0) grid (2,2);
		\end{scope}		
		
		\begin{scope}[shift={(2,0)}]
			\fill[red!50!black] (1,0) rectangle (2,1);
			\fill[blue!50!black] (0,1) rectangle (1,2);
			
			\draw[step=1cm,black,very thin] (0,0) grid (2,2);
		\end{scope}		
		
		\draw[step=2cm,black,very thin] (0,0) grid (4,4);

	\end{scope}

	\begin{scope}[shift={(5,-5)}]

		\fill[red!50!black] (0,2) rectangle (1,3);
		\fill[blue!50!black] (1,3) rectangle (2,4);
		
		\draw[step=1cm,black,very thin] (0,2) grid (2,4);
		
		\begin{scope}[shift={(2,-2)}]
			\fill[red!50!black] (0,2) rectangle (1,3);
			\fill[blue!50!black] (1,3) rectangle (2,4);
			
			\draw[step=1cm,black,very thin] (0,2) grid (2,4);
		\end{scope}		
		
		\draw[step=2cm,black,very thin] (0,0) grid (4,4);

	\end{scope}

	\begin{scope}[shift={(10,-10)}]
		
		\fill[orange!75!black] (0,2) rectangle (1,3);
		\fill[orange!75!black] (1,3) rectangle (2,4);

		\draw[step=1cm,black,very thin] (0,2) grid (2,4);
		
		\begin{scope}[shift={(2,-2)}]
			\fill[green!50!black] (0,2) rectangle (1,3);
			\fill[green!50!black] (1,3) rectangle (2,4);
			\draw[step=1cm,black,very thin] (0,2) grid (2,4);
		\end{scope}		
		
		\draw[step=2cm,black,very thin] (0,0) grid (4,4);

	\end{scope}

	\begin{scope}[shift={(10,-15)}]

		\begin{scope}[shift={(2,2)}]
			\fill[blue!50!black] (0,0) rectangle (1,1);
			\fill[blue!50!black] (1,1) rectangle (2,2);
			
			\draw[step=1cm,black,very thin] (0,0) grid (2,2);
		\end{scope}		
		
		\begin{scope}
			\fill[red!50!black] (0,0) rectangle (1,1);
			\fill[red!50!black] (1,1) rectangle (2,2);
			
			\draw[step=1cm,black,very thin] (0,0) grid (2,2);
		\end{scope}

		\draw[step=2cm,black,very thin] (0,0) grid (4,4);
		
	\end{scope}

\end{scope}}}
		\end{aligned}\,.
	\end{align}
	Therefore, the 4-dimensional set, complete of its chiral matrix, is given by
	\begin{align}\label{eq:4dimVerbo}
		\begin{aligned}
			\gamma_{(4)}^0&=\id_2\otimes\,\sigma^3
			\,,
			\\
			\gamma_{(4)}^1&=\id_2\otimes\,\iu\sigma^2
			\,,
			\\
			\gamma_{(4)}^2&=-\iu\sigma^3\otimes\sigma^1
			\,,
		\end{aligned}
		&&
		\begin{aligned}
			\gamma_{(4)}^3&=\iu\sigma^2\otimes\sigma^1
			\,,
			\\
			\gamma_{(4)}^{\rm ch}&=\sigma^1\otimes \sigma^1
			\,,
		\end{aligned}
	\end{align}
	i.e., by putting the gamma matrices inside the box and the chiral matrix outside,
	\begin{align}\label{eq:4dim}
		\begin{aligned}
			\raisebox{-1.5pt}{\tikz{		\begin{scope}[scale=.36]


			\draw [
ultra thick,orange!70!black!50,rounded corners=1pt]  (-.3,4.3) -- (-.3,-.3) --(19.3,-.3) --(19.3,4.3) --cycle;


	
	\begin{scope}[shift={(0,2)}]
		\fill[red!50!black] (1,0) rectangle (2,1);
		\fill[blue!50!black] (0,1) rectangle (1,2);
		
		\draw[step=1cm,black,very thin] (0,0) grid (2,2);
	\end{scope}		
	
	\begin{scope}[shift={(2,0)}]
		\fill[red!50!black] (1,0) rectangle (2,1);
		\fill[blue!50!black] (0,1) rectangle (1,2);
		
		\draw[step=1cm,black,very thin] (0,0) grid (2,2);
	\end{scope}		
	
	\draw[step=2cm,black,very thin] (0,0) grid (4,4);

	\begin{scope}[shift={(5,0)}]

		\fill[red!50!black] (0,2) rectangle (1,3);
		\fill[blue!50!black] (1,3) rectangle (2,4);
		
		\draw[step=1cm,black,very thin] (0,2) grid (2,4);
		
		\begin{scope}[shift={(2,-2)}]
			\fill[red!50!black] (0,2) rectangle (1,3);
			\fill[blue!50!black] (1,3) rectangle (2,4);
			
			\draw[step=1cm,black,very thin] (0,2) grid (2,4);
		\end{scope}		
		
		\draw[step=2cm,black,very thin] (0,0) grid (4,4);

	\end{scope}

	\begin{scope}[shift={(10,0)}]
		
		\fill[orange!75!black] (0,2) rectangle (1,3);
		\fill[orange!75!black] (1,3) rectangle (2,4);

		\draw[step=1cm,black,very thin] (0,2) grid (2,4);
		
		\begin{scope}[shift={(2,-2)}]
			\fill[green!50!black] (0,2) rectangle (1,3);
			\fill[green!50!black] (1,3) rectangle (2,4);
			\draw[step=1cm,black,very thin] (0,2) grid (2,4);
		\end{scope}		
		
		\draw[step=2cm,black,very thin] (0,0) grid (4,4);		
		
	\end{scope}

	\begin{scope}[shift={(15,0)}]
		
		\begin{scope}[shift={(2,2)}]
			\fill[blue!50!black] (0,0) rectangle (1,1);
			\fill[blue!50!black] (1,1) rectangle (2,2);
			
			\draw[step=1cm,black,very thin] (0,0) grid (2,2);
		\end{scope}		
		
		\begin{scope}
			\fill[red!50!black] (0,0) rectangle (1,1);
			\fill[red!50!black] (1,1) rectangle (2,2);
			
			\draw[step=1cm,black,very thin] (0,0) grid (2,2);
		\end{scope}

		\draw[step=2cm,black,very thin] (0,0) grid (4,4);

	\end{scope}

	\begin{scope}[shift={(20,0)}]

		\begin{scope}[shift={(2,2)}]
			\fill[blue!50!black] (0,0) rectangle (1,1);
			\fill[blue!50!black] (1,1) rectangle (2,2);
			
			\draw[step=1cm,black,very thin] (0,0) grid (2,2);
		\end{scope}		
		
		\begin{scope}
			\fill[blue!50!black] (0,0) rectangle (1,1);
			\fill[blue!50!black] (1,1) rectangle (2,2);
			
			\draw[step=1cm,black,very thin] (0,0) grid (2,2);
		\end{scope}

		\draw[step=2cm,black,very thin] (0,0) grid (4,4);
		
	\end{scope}

\end{scope}}}
		\end{aligned}\,.
	\end{align}			
	
	\paragraph{}
	\noindent{\bf The 5-dimensional set} is just the 4-dimensional plus $\iu$ times its chiral matrix,
	\begin{align}\label{eq:5dimVerbo}
		\begin{aligned}
			\gamma_{(5)}^0&=\id_2\otimes\sigma^3
			\,,
			\\
			\gamma_{(5)}^1&=\id_2\otimes\,\iu\sigma^2
			\,,
			\\
			\gamma_{(5)}^2&=-\iu\sigma^3\otimes\sigma^1
			\,,
		\end{aligned}
		&&
		\begin{aligned}
			\gamma_{(5)}^3&=\iu\sigma^2\otimes\sigma^1
			\,,
			\\
			\gamma_{(5)}^{4}&=\iu\sigma^1\otimes \sigma^1
			\,,
		\end{aligned}
	\end{align}
	that is, we take the set~\eqref{eq:4dim} and multiply the matrix left out of the contour by $-\iu$,
	\begin{align}\label{eq:5dim}
		\begin{aligned}
			\raisebox{-1.5pt}{\tikz{\begin{scope}[scale=.36]

	
	\begin{scope}[shift={(0,2)}]
		\fill[red!50!black] (1,0) rectangle (2,1);
		\fill[blue!50!black] (0,1) rectangle (1,2);
		
		\draw[step=1cm,black,very thin] (0,0) grid (2,2);
	\end{scope}		
	
	\begin{scope}[shift={(2,0)}]
		\fill[red!50!black] (1,0) rectangle (2,1);
		\fill[blue!50!black] (0,1) rectangle (1,2);
		
		\draw[step=1cm,black,very thin] (0,0) grid (2,2);
	\end{scope}		
	
	\draw[step=2cm,black,very thin] (0,0) grid (4,4);

	\begin{scope}[shift={(5,0)}]

		\fill[red!50!black] (0,2) rectangle (1,3);
		\fill[blue!50!black] (1,3) rectangle (2,4);
		
		\draw[step=1cm,black,very thin] (0,2) grid (2,4);
		
		\begin{scope}[shift={(2,-2)}]
			\fill[red!50!black] (0,2) rectangle (1,3);
			\fill[blue!50!black] (1,3) rectangle (2,4);
			
			\draw[step=1cm,black,very thin] (0,2) grid (2,4);
		\end{scope}		
		
		\draw[step=2cm,black,very thin] (0,0) grid (4,4);

	\end{scope}

	\begin{scope}[shift={(10,0)}]
		
		\fill[orange!75!black] (0,2) rectangle (1,3);
		\fill[orange!75!black] (1,3) rectangle (2,4);

		\draw[step=1cm,black,very thin] (0,2) grid (2,4);
		
		\begin{scope}[shift={(2,-2)}]
			\fill[green!50!black] (0,2) rectangle (1,3);
			\fill[green!50!black] (1,3) rectangle (2,4);
			\draw[step=1cm,black,very thin] (0,2) grid (2,4);
		\end{scope}		
		
		\draw[step=2cm,black,very thin] (0,0) grid (4,4);		
		
	\end{scope}

	\begin{scope}[shift={(15,0)}]
		
		\begin{scope}[shift={(2,2)}]
			\fill[blue!50!black] (0,0) rectangle (1,1);
			\fill[blue!50!black] (1,1) rectangle (2,2);
			
			\draw[step=1cm,black,very thin] (0,0) grid (2,2);
		\end{scope}		
		
		\begin{scope}
			\fill[red!50!black] (0,0) rectangle (1,1);
			\fill[red!50!black] (1,1) rectangle (2,2);
			
			\draw[step=1cm,black,very thin] (0,0) grid (2,2);
		\end{scope}

		\draw[step=2cm,black,very thin] (0,0) grid (4,4);

	\end{scope}

	\begin{scope}[shift={(20,0)}]

		\begin{scope}[shift={(2,2)}]
			\fill[orange!75!black] (0,0) rectangle (1,1);
			\fill[orange!75!black] (1,1) rectangle (2,2);
			
			\draw[step=1cm,black,very thin] (0,0) grid (2,2);
		\end{scope}		
		
		\begin{scope}
			\fill[orange!75!black] (0,0) rectangle (1,1);
			\fill[orange!75!black] (1,1) rectangle (2,2);
			
			\draw[step=1cm,black,very thin] (0,0) grid (2,2);
		\end{scope}

		\draw[step=2cm,black,very thin] (0,0) grid (4,4);
		
	\end{scope}

\end{scope}}}
		\end{aligned}\,,
	\end{align}
	
	\paragraph{}
	\noindent{\bf The 6-dimensional set} is made up of six Kronecker products: the first four matrices come from the products between $\id_2$ and the matrices in 4 dimensions, the last two from the products between the fundamental matrices $-\iu\sigma^3$ and $\iu\sigma^2$ and the chiral matrix in 4 dimensions. Therefore, we consider the $3\times5$ multiplicative table of the Kronecker product, with column entries~\eqref{eq:left}, and row entries~\eqref{eq:2dim}. As before, we only need the products in the first row and in the last column, without the one at their intersection:
	\begin{align}
		\begin{aligned}
			\raisebox{-1.5pt}{\tikz{		\begin{scope}[scale=.25]

	
	\draw[very thin,fill=gray!20!white] (-0.5,-0.5) -- (44.5,-0.5) [rounded corners]-- (44.5,4.5) -- (-0.5,4.5) [sharp corners]-- cycle;
	\draw[very thin,fill=gray!20!white] (-0.5,-0.5) -- (-0.5,-27.5) [rounded corners]-- (-9.5,-27.5) -- (-9.5,-.5) [sharp corners]-- cycle;

	\begin{scope}[shift={(2,0)}]			
		
		\draw [
	ultra thick,orange!70!black!50,rounded corners=1pt] (-.3,-.3) -- (-.3,4.3) --(31.3,4.3) --(31.3,-.3) --cycle;	
		\end{scope}	
	
	\draw[help lines,step=9,xshift=-.5cm,yshift=-.5cm] (0,0) grid (45,-27);

	\begin{scope}[shift={(2,0)}]

		
		\fill[white] (0,0) rectangle (4,4);				
		
		\begin{scope}[shift={(0,2)}]

			\fill[red!50!black] (1,0) rectangle (2,1);
			\fill[blue!50!black] (0,1) rectangle (1,2);
			
			\draw[step=1cm,black,very thin] (0,0) grid (2,2);
		\end{scope}		
		
		\begin{scope}[shift={(2,0)}]
			\fill[red!50!black] (1,0) rectangle (2,1);
			\fill[blue!50!black] (0,1) rectangle (1,2);
			
			\draw[step=1cm,black,very thin] (0,0) grid (2,2);
		\end{scope}		
		
		\draw[step=2cm,black,very thin] (0,0) grid (4,4);

		\begin{scope}[shift={(9,0)}]
			
			\fill[white] (0,0) rectangle (4,4);				
			
			\fill[red!50!black] (0,2) rectangle (1,3);
			\fill[blue!50!black] (1,3) rectangle (2,4);
			
			\draw[step=1cm,black,very thin] (0,2) grid (2,4);
			
			\begin{scope}[shift={(2,-2)}]
				\fill[red!50!black] (0,2) rectangle (1,3);
				\fill[blue!50!black] (1,3) rectangle (2,4);
				
				\draw[step=1cm,black,very thin] (0,2) grid (2,4);
			\end{scope}		
			
			\draw[step=2cm,black,very thin] (0,0) grid (4,4);

		\end{scope}

		\begin{scope}[shift={(18,0)}]
			
			\fill[white] (0,0) rectangle (4,4);				
			
			\fill[orange!75!black] (0,2) rectangle (1,3);
			\fill[orange!75!black] (1,3) rectangle (2,4);

			\draw[step=1cm,black,very thin] (0,2) grid (2,4);
			
			\begin{scope}[shift={(2,-2)}]
				\fill[green!50!black] (0,2) rectangle (1,3);
				\fill[green!50!black] (1,3) rectangle (2,4);
				\draw[step=1cm,black,very thin] (0,2) grid (2,4);
			\end{scope}		
			
			\draw[step=2cm,black,very thin] (0,0) grid (4,4);		
			
		\end{scope}

		\begin{scope}[shift={(27,0)}]
			
			\fill[white] (0,0) rectangle (4,4);				
			
			\begin{scope}[shift={(2,2)}]
				\fill[blue!50!black] (0,0) rectangle (1,1);
				\fill[blue!50!black] (1,1) rectangle (2,2);
				
				\draw[step=1cm,black,very thin] (0,0) grid (2,2);
			\end{scope}		
			
			\begin{scope}
				\fill[red!50!black] (0,0) rectangle (1,1);
				\fill[red!50!black] (1,1) rectangle (2,2);
				
				\draw[step=1cm,black,very thin] (0,0) grid (2,2);
			\end{scope}

			\draw[step=2cm,black,very thin] (0,0) grid (4,4);

		\end{scope}	
		
		\begin{scope}[shift={(36,0)}]
			
			\fill[white] (0,0) rectangle (4,4);				
			
			\begin{scope}[shift={(2,2)}]
				\fill[blue!50!black] (0,0) rectangle (1,1);
				\fill[blue!50!black] (1,1) rectangle (2,2);
				
				\draw[step=1cm,black,very thin] (0,0) grid (2,2);
			\end{scope}		
			
			\begin{scope}
				\fill[blue!50!black] (0,0) rectangle (1,1);
				\fill[blue!50!black] (1,1) rectangle (2,2);
				
				\draw[step=1cm,black,very thin] (0,0) grid (2,2);
			\end{scope}

			\draw[step=2cm,black,very thin] (0,0) grid (4,4);
			
		\end{scope}

	\end{scope}


	\begin{scope}[shift={(-12,-4)}]
		\draw [	ultra thick,cyan!70!black!50,rounded corners=1pt]  (-.3,.3) -- (2.3,.3) --(2.3,-20.3) --(-.3,-20.3) --cycle;
	\end{scope}

	\begin{scope}[shift={(-9,-9)}]
		
		\fill[white] (0,0) rectangle (8,8);
		
		\draw[step=4cm,black,very thin] (0,0) grid (8,8);
		
		\draw [fill=blue!50!black!,
		draw=blue!50!black!, thick,rounded corners=1pt,fill opacity=0.15](0,4) rectangle (4,8);
		
		\draw [fill=blue!50!black!,
		draw=blue!50!black!, thick,rounded corners=1pt,fill opacity=0.15](4,0) rectangle (8,4);	
		
	\end{scope}

	\begin{scope}[shift={(-9,-18)}]

		\fill[white] (0,0) rectangle (8,8);
		
		\draw[step=4cm,black,very thin] (0,0) grid (8,8);
		
		\draw [fill=orange!75!black!,
		draw=orange!75!black!, thick,rounded corners=1pt,fill opacity=0.15](0,4) rectangle (4,8);
		
		\draw [fill=green!50!black!,
		draw=green!50!black!, thick,rounded corners=1pt,fill opacity=0.15](4,0) rectangle (8,4);	
		
	\end{scope}

	\begin{scope}[shift={(-9,-27)}]
		
		\fill[white] (0,0) rectangle (8,8);
		
		\draw[step=4cm,black,very thin] (0,0) grid (8,8);
		
		\draw [fill=blue!50!black!,
		draw=blue!50!black!, thick,rounded corners=1pt,fill opacity=0.15](4,4) rectangle (8,8);
		
		\draw [fill=red!50!black!,
		draw=red!50!black!, thick,rounded corners=1pt,fill opacity=0.15](0,0) rectangle (4,4);	
		
	\end{scope}

	\begin{scope}[shift={(-12,-6)}]
		
		\fill[blue!50!black] (1,0) rectangle (2,1);
		\fill[blue!50!black] (0,1) rectangle (1,2);
		
		\draw[step=1cm,black,very thin] (0,0) grid (2,2);
		
		\draw[ultra thin, gray] (0,2) -- (3,5);	
		\draw[ultra thin, gray] (2,2) -- (11,5);				
		\draw[ultra thin, gray] (2,0) -- (11,-3);				
		\draw[ultra thin, gray] (0,0) -- (3,-3);

	\end{scope}

	\begin{scope}[shift={(-12,-15)}]
		
		\fill[orange!75!black] (0,1) rectangle (1,2);
		\fill[green!50!black] (1,0) rectangle (2,1);
		
		\draw[step=1cm,black,very thin] (0,0) grid (2,2);
		
		\draw[ultra thin, gray] (0,2) -- (3,5);	
		\draw[ultra thin, gray] (2,2) -- (11,5);				
		\draw[ultra thin, gray] (2,0) -- (11,-3);				
		\draw[ultra thin, gray] (0,0) -- (3,-3);

	\end{scope}		
	
	\begin{scope}[shift={(-12,-24)}]
		
		\fill[red!50!black] (0,0) rectangle (1,1);
		\fill[blue!50!black] (1,1) rectangle (2,2);
		
		\draw[step=1cm,black,very thin] (0,0) grid (2,2);
		
		\draw[ultra thin, gray] (0,2) -- (3,5);	
		\draw[ultra thin, gray] (2,2) -- (11,5);				
		\draw[ultra thin, gray] (2,0) -- (11,-3);				
		\draw[ultra thin, gray] (0,0) -- (3,-3);		
		
	\end{scope}

	
	\begin{scope}[shift={(0,-9)}]
		
		\begin{scope}[shift={(0,4)}]
			
			\begin{scope}[shift={(0,2)}]
				\fill[red!50!black] (1,0) rectangle (2,1);
				\fill[blue!50!black] (0,1) rectangle (1,2);
				\draw[step=1cm,black,very thin] (0,0) grid (2,2);	
			\end{scope}		
			
			\begin{scope}[shift={(2,0)}]
				\fill[red!50!black] (1,0) rectangle (2,1);
				\fill[blue!50!black] (0,1) rectangle (1,2);
				\draw[step=1cm,black,very thin] (0,0) grid (2,2);	
			\end{scope}		
			\draw[step=2cm,black,very thin] (0,0) grid (4,4);
		\end{scope}	
		
		\begin{scope}[shift={(4,0)}]

			\begin{scope}[shift={(0,2)}]
				\fill[red!50!black] (1,0) rectangle (2,1);
				\fill[blue!50!black] (0,1) rectangle (1,2);
				\draw[step=1cm,black,very thin] (0,0) grid (2,2);	
			\end{scope}		
			
			\begin{scope}[shift={(2,0)}]
				\fill[red!50!black] (1,0) rectangle (2,1);
				\fill[blue!50!black] (0,1) rectangle (1,2);
				\draw[step=1cm,black,very thin] (0,0) grid (2,2);	
			\end{scope}		
			\draw[step=2cm,black,very thin] (0,0) grid (4,4);
		\end{scope}	
		
		\draw[step=4cm,black,very thin] (0,0) grid (8,8);
	\end{scope}

	\begin{scope}[shift={(9,-9)}]
		
		\begin{scope}[shift={(0,4)}]
			
			\begin{scope}[shift={(0,2)}]
				\fill[red!50!black] (0,0) rectangle (1,1);
				\fill[blue!50!black] (1,1) rectangle (2,2);
				\draw[step=1cm,black,very thin] (0,0) grid (2,2);	
			\end{scope}		
			
			\begin{scope}[shift={(2,0)}]
				\fill[red!50!black] (0,0) rectangle (1,1);
				\fill[blue!50!black] (1,1) rectangle (2,2);
				\draw[step=1cm,black,very thin] (0,0) grid (2,2);	
			\end{scope}		
			\draw[step=2cm,black,very thin] (0,0) grid (4,4);
		\end{scope}	
		
		\begin{scope}[shift={(4,0)}]

			\begin{scope}[shift={(0,2)}]
				\fill[red!50!black] (0,0) rectangle (1,1);
				\fill[blue!50!black] (1,1) rectangle (2,2);
				\draw[step=1cm,black,very thin] (0,0) grid (2,2);	
			\end{scope}		
			
			\begin{scope}[shift={(2,0)}]
				\fill[red!50!black] (0,0) rectangle (1,1);
				\fill[blue!50!black] (1,1) rectangle (2,2);
				\draw[step=1cm,black,very thin] (0,0) grid (2,2);	
			\end{scope}		
			\draw[step=2cm,black,very thin] (0,0) grid (4,4);
		\end{scope}	
		
		\draw[step=4cm,black,very thin] (0,0) grid (8,8);
	\end{scope}

	\begin{scope}[shift={(18,-9)}]
		
		\begin{scope}[shift={(0,4)}]
			
			\fill[orange!75!black] (0,2) rectangle (1,3);
			\fill[orange!75!black] (1,3) rectangle (2,4);

			\draw[step=1cm,black,very thin] (0,2) grid (2,4);
			
			\begin{scope}[shift={(2,-2)}]
				\fill[green!50!black] (0,2) rectangle (1,3);
				\fill[green!50!black] (1,3) rectangle (2,4);
				\draw[step=1cm,black,very thin] (0,2) grid (2,4);
			\end{scope}		
			
			\draw[step=2cm,black,very thin] (0,0) grid (4,4);		
			
		\end{scope}	
		
		\begin{scope}[shift={(4,0)}]

			\fill[orange!75!black] (0,2) rectangle (1,3);
			\fill[orange!75!black] (1,3) rectangle (2,4);

			\draw[step=1cm,black,very thin] (0,2) grid (2,4);
			
			\begin{scope}[shift={(2,-2)}]
				\fill[green!50!black] (0,2) rectangle (1,3);
				\fill[green!50!black] (1,3) rectangle (2,4);
				\draw[step=1cm,black,very thin] (0,2) grid (2,4);
			\end{scope}		
			
			\draw[step=2cm,black,very thin] (0,0) grid (4,4);		
		\end{scope}	
		
		\draw[step=4cm,black,very thin] (0,0) grid (8,8);
	\end{scope}

	
	\begin{scope}[shift={(27,-9)}]
		
		\begin{scope}[shift={(0,4)}]

			\begin{scope}[shift={(2,2)}]
				\fill[blue!50!black] (0,0) rectangle (1,1);
				\fill[blue!50!black] (1,1) rectangle (2,2);
				
				\draw[step=1cm,black,very thin] (0,0) grid (2,2);
			\end{scope}		
			
			\begin{scope}
				\fill[red!50!black] (0,0) rectangle (1,1);
				\fill[red!50!black] (1,1) rectangle (2,2);
				
				\draw[step=1cm,black,very thin] (0,0) grid (2,2);
			\end{scope}

			\draw[step=2cm,black,very thin] (0,0) grid (4,4);
			
		\end{scope}	
		
		\begin{scope}[shift={(4,0)}]

			\begin{scope}[shift={(2,2)}]
				\fill[blue!50!black] (0,0) rectangle (1,1);
				\fill[blue!50!black] (1,1) rectangle (2,2);
				
				\draw[step=1cm,black,very thin] (0,0) grid (2,2);
			\end{scope}		
			
			\begin{scope}
				\fill[red!50!black] (0,0) rectangle (1,1);
				\fill[red!50!black] (1,1) rectangle (2,2);
				
				\draw[step=1cm,black,very thin] (0,0) grid (2,2);
			\end{scope}

			\draw[step=2cm,black,very thin] (0,0) grid (4,4);
		\end{scope}	
		
		\draw[step=4cm,black,very thin] (0,0) grid (8,8);
	\end{scope}

	\begin{scope}[shift={(36,-18)}]

		\begin{scope}[shift={(0,4)}]

			\fill[orange!75!black] (0,0) rectangle (1,1);
			\fill[orange!75!black] (1,1) rectangle (2,2);

			\draw[step=1cm,black,very thin] (0,0) grid (2,2);
			
			\begin{scope}[shift={(2,2)}]
				\fill[orange!75!black] (0,0) rectangle (1,1);
				\fill[orange!75!black] (1,1) rectangle (2,2);
				
				\draw[step=1cm,black,very thin] (0,0) grid (2,2);
				
			\end{scope}

			\draw[step=2cm,black,very thin] (0,0) grid (4,4);
		\end{scope}	
		
		\begin{scope}[shift={(4,0)}]

			\fill[green!50!black] (0,0) rectangle (1,1);
			\fill[green!50!black] (1,1) rectangle (2,2);

			\draw[step=1cm,black,very thin] (0,0) grid (2,2);
			
			\begin{scope}[shift={(2,2)}]
				\fill[green!50!black] (0,0) rectangle (1,1);
				\fill[green!50!black] (1,1) rectangle (2,2);
				
				\draw[step=1cm,black,very thin] (0,0) grid (2,2);
				
			\end{scope}		
			
			\draw[step=2cm,black,very thin] (0,0) grid (4,4);
			
		\end{scope}			
		
		\draw[step=4cm,black,very thin] (0,0) grid (8,8);
	\end{scope}

	
	\begin{scope}[shift={(36,-27)}]
		\draw[step=4cm,black,very thin] (0,0) grid (8,8);
		
		\begin{scope}[shift={(4,4)}]
			
			\begin{scope}[shift={(2,2)}]
				\fill[blue!50!black] (0,0) rectangle (1,1);
				\fill[blue!50!black] (1,1) rectangle (2,2);
				
				\draw[step=1cm,black,very thin] (0,0) grid (2,2);
			\end{scope}		
			
			\begin{scope}
				\fill[blue!50!black] (0,0) rectangle (1,1);
				\fill[blue!50!black] (1,1) rectangle (2,2);
				
				\draw[step=1cm,black,very thin] (0,0) grid (2,2);
			\end{scope}

			\draw[step=2cm,black,very thin] (0,0) grid (4,4);
		\end{scope}

		\begin{scope}[shift={(0,0)}]
			
			\begin{scope}[shift={(2,2)}]
				\fill[red!50!black] (0,0) rectangle (1,1);
				\fill[red!50!black] (1,1) rectangle (2,2);
				
				\draw[step=1cm,black,very thin] (0,0) grid (2,2);
			\end{scope}		
			
			\begin{scope}
				\fill[red!50!black] (0,0) rectangle (1,1);
				\fill[red!50!black] (1,1) rectangle (2,2);
				
				\draw[step=1cm,black,very thin] (0,0) grid (2,2);
			\end{scope}		
			
			\draw[step=2cm,black,very thin] (0,0) grid (4,4);
		\end{scope}

	\end{scope}

\end{scope}}}
		\end{aligned}
		\,.
	\end{align}
	Finally, the 6-dimensional set (cf.\ \cite{Littlewood}), complete of its chiral matrix, is given by
	\begin{align}\label{eq:6dimVerbo}
		\begin{aligned}
			\gamma_{(6)}^0&=\id_2\otimes\id_2\otimes\,\sigma^3
			\,,
			\\
			\gamma_{(6)}^1&=\iu\id_2\otimes\id_2\otimes\,\sigma^2
			\,,
			\\
			\gamma_{(6)}^2&=-\iu\id_2\otimes\,\sigma^3\otimes\sigma^1
			\,,
			\\
			\gamma_{(6)}^3&=\iu\id_2\otimes\sigma^2\otimes\sigma^1
			\,,
		\end{aligned}
		&&
		\begin{aligned}
			\gamma_{(6)}^4&=-\iu\sigma^3\otimes\sigma^1\otimes\sigma^1
			\,,
			\\
			\gamma_{(6)}^5&=\iu\sigma^2\otimes\sigma^1\otimes\sigma^1
			\,,
			\\
			\gamma_{(6)}^{\rm ch}&=\sigma^1\otimes \sigma^1\otimes\sigma^1
			\,,
		\end{aligned}
	\end{align}					
	namely, by putting the gamma matrices inside the box and the chiral matrix outside,
	\begin{align}\label{eq:6dim}
		\begin{aligned}
			\raisebox{-1.5pt}{\tikz{	\begin{scope}[scale=.23]

	\draw [
ultra thick,green!30!black!50,rounded corners=1pt]  (-.3,8.3) -- (-.3,-.3) --(53.3,-.3) --(53.3,8.3) --cycle;


	\begin{scope}[shift={(0,0)}]
		
		\begin{scope}[shift={(0,4)}]
			
			\begin{scope}[shift={(0,2)}]
				\fill[red!50!black] (1,0) rectangle (2,1);
				\fill[blue!50!black] (0,1) rectangle (1,2);
				\draw[step=1cm,black,very thin] (0,0) grid (2,2);	
			\end{scope}		
			
			\begin{scope}[shift={(2,0)}]
				\fill[red!50!black] (1,0) rectangle (2,1);
				\fill[blue!50!black] (0,1) rectangle (1,2);
				\draw[step=1cm,black,very thin] (0,0) grid (2,2);	
			\end{scope}		
			\draw[step=2cm,black,very thin] (0,0) grid (4,4);
		\end{scope}	
		
		\begin{scope}[shift={(4,0)}]

			\begin{scope}[shift={(0,2)}]
				\fill[red!50!black] (1,0) rectangle (2,1);
				\fill[blue!50!black] (0,1) rectangle (1,2);
				\draw[step=1cm,black,very thin] (0,0) grid (2,2);	
			\end{scope}		
			
			\begin{scope}[shift={(2,0)}]
				\fill[red!50!black] (1,0) rectangle (2,1);
				\fill[blue!50!black] (0,1) rectangle (1,2);
				\draw[step=1cm,black,very thin] (0,0) grid (2,2);	
			\end{scope}		
			\draw[step=2cm,black,very thin] (0,0) grid (4,4);
		\end{scope}	
		
		\draw[step=4cm,black,very thin] (0,0) grid (8,8);
	\end{scope}

	\begin{scope}[shift={(9,0)}]
		
		\begin{scope}[shift={(0,4)}]
			
			\begin{scope}[shift={(0,2)}]
				\fill[red!50!black] (0,0) rectangle (1,1);
				\fill[blue!50!black] (1,1) rectangle (2,2);
				\draw[step=1cm,black,very thin] (0,0) grid (2,2);	
			\end{scope}		
			
			\begin{scope}[shift={(2,0)}]
				\fill[red!50!black] (0,0) rectangle (1,1);
				\fill[blue!50!black] (1,1) rectangle (2,2);
				\draw[step=1cm,black,very thin] (0,0) grid (2,2);	
			\end{scope}		
			\draw[step=2cm,black,very thin] (0,0) grid (4,4);
		\end{scope}	
		
		\begin{scope}[shift={(4,0)}]

			\begin{scope}[shift={(0,2)}]
				\fill[red!50!black] (0,0) rectangle (1,1);
				\fill[blue!50!black] (1,1) rectangle (2,2);
				\draw[step=1cm,black,very thin] (0,0) grid (2,2);	
			\end{scope}		
			
			\begin{scope}[shift={(2,0)}]
				\fill[red!50!black] (0,0) rectangle (1,1);
				\fill[blue!50!black] (1,1) rectangle (2,2);
				\draw[step=1cm,black,very thin] (0,0) grid (2,2);	
			\end{scope}		
			\draw[step=2cm,black,very thin] (0,0) grid (4,4);
		\end{scope}	
		
		\draw[step=4cm,black,very thin] (0,0) grid (8,8);
	\end{scope}

	\begin{scope}[shift={(18,0)}]
		
		\begin{scope}[shift={(0,4)}]
			
			\fill[orange!75!black] (0,2) rectangle (1,3);
			\fill[orange!75!black] (1,3) rectangle (2,4);

			\draw[step=1cm,black,very thin] (0,2) grid (2,4);
			
			\begin{scope}[shift={(2,-2)}]
				\fill[green!50!black] (0,2) rectangle (1,3);
				\fill[green!50!black] (1,3) rectangle (2,4);
				\draw[step=1cm,black,very thin] (0,2) grid (2,4);
			\end{scope}		
			
			\draw[step=2cm,black,very thin] (0,0) grid (4,4);		
			
		\end{scope}	
		
		\begin{scope}[shift={(4,0)}]

			\fill[orange!75!black] (0,2) rectangle (1,3);
			\fill[orange!75!black] (1,3) rectangle (2,4);

			\draw[step=1cm,black,very thin] (0,2) grid (2,4);
			
			\begin{scope}[shift={(2,-2)}]
				\fill[green!50!black] (0,2) rectangle (1,3);
				\fill[green!50!black] (1,3) rectangle (2,4);
				\draw[step=1cm,black,very thin] (0,2) grid (2,4);
			\end{scope}		
			
			\draw[step=2cm,black,very thin] (0,0) grid (4,4);		
		\end{scope}	
		
		\draw[step=4cm,black,very thin] (0,0) grid (8,8);
	\end{scope}

	
	\begin{scope}[shift={(27,0)}]
		
		\begin{scope}[shift={(0,4)}]

			\begin{scope}[shift={(2,2)}]
				\fill[blue!50!black] (0,0) rectangle (1,1);
				\fill[blue!50!black] (1,1) rectangle (2,2);
				
				\draw[step=1cm,black,very thin] (0,0) grid (2,2);
			\end{scope}		
			
			\begin{scope}
				\fill[red!50!black] (0,0) rectangle (1,1);
				\fill[red!50!black] (1,1) rectangle (2,2);
				
				\draw[step=1cm,black,very thin] (0,0) grid (2,2);
			\end{scope}

			\draw[step=2cm,black,very thin] (0,0) grid (4,4);
			
		\end{scope}	
		
		\begin{scope}[shift={(4,0)}]

			\begin{scope}[shift={(2,2)}]
				\fill[blue!50!black] (0,0) rectangle (1,1);
				\fill[blue!50!black] (1,1) rectangle (2,2);
				
				\draw[step=1cm,black,very thin] (0,0) grid (2,2);
			\end{scope}		
			
			\begin{scope}
				\fill[red!50!black] (0,0) rectangle (1,1);
				\fill[red!50!black] (1,1) rectangle (2,2);
				
				\draw[step=1cm,black,very thin] (0,0) grid (2,2);
			\end{scope}

			\draw[step=2cm,black,very thin] (0,0) grid (4,4);
		\end{scope}	
		
		\draw[step=4cm,black,very thin] (0,0) grid (8,8);
	\end{scope}

	\begin{scope}[shift={(36,0)}]

		\begin{scope}[shift={(0,4)}]

			\fill[orange!75!black] (0,0) rectangle (1,1);
			\fill[orange!75!black] (1,1) rectangle (2,2);

			\draw[step=1cm,black,very thin] (0,0) grid (2,2);
			
			\begin{scope}[shift={(2,2)}]
				\fill[orange!75!black] (0,0) rectangle (1,1);
				\fill[orange!75!black] (1,1) rectangle (2,2);
				
				\draw[step=1cm,black,very thin] (0,0) grid (2,2);
				
			\end{scope}

			\draw[step=2cm,black,very thin] (0,0) grid (4,4);
		\end{scope}	
		
		\begin{scope}[shift={(4,0)}]

			\fill[green!50!black] (0,0) rectangle (1,1);
			\fill[green!50!black] (1,1) rectangle (2,2);

			\draw[step=1cm,black,very thin] (0,0) grid (2,2);
			
			\begin{scope}[shift={(2,2)}]
				\fill[green!50!black] (0,0) rectangle (1,1);
				\fill[green!50!black] (1,1) rectangle (2,2);
				
				\draw[step=1cm,black,very thin] (0,0) grid (2,2);
				
			\end{scope}		
			
			\draw[step=2cm,black,very thin] (0,0) grid (4,4);
			
		\end{scope}			
		
		\draw[step=4cm,black,very thin] (0,0) grid (8,8);
	\end{scope}

	
	\begin{scope}[shift={(45,0)}]
		\draw[step=4cm,black,very thin] (0,0) grid (8,8);
		
		\begin{scope}[shift={(4,4)}]
			
			\begin{scope}[shift={(2,2)}]
				\fill[blue!50!black] (0,0) rectangle (1,1);
				\fill[blue!50!black] (1,1) rectangle (2,2);
				
				\draw[step=1cm,black,very thin] (0,0) grid (2,2);
			\end{scope}		
			
			\begin{scope}
				\fill[blue!50!black] (0,0) rectangle (1,1);
				\fill[blue!50!black] (1,1) rectangle (2,2);
				
				\draw[step=1cm,black,very thin] (0,0) grid (2,2);
			\end{scope}

			\draw[step=2cm,black,very thin] (0,0) grid (4,4);
		\end{scope}

		\begin{scope}[shift={(0,0)}]
			
			\begin{scope}[shift={(2,2)}]
				\fill[red!50!black] (0,0) rectangle (1,1);
				\fill[red!50!black] (1,1) rectangle (2,2);
				
				\draw[step=1cm,black,very thin] (0,0) grid (2,2);
			\end{scope}		
			
			\begin{scope}
				\fill[red!50!black] (0,0) rectangle (1,1);
				\fill[red!50!black] (1,1) rectangle (2,2);
				
				\draw[step=1cm,black,very thin] (0,0) grid (2,2);
			\end{scope}		
			
			\draw[step=2cm,black,very thin] (0,0) grid (4,4);
		\end{scope}

	\end{scope}

	\begin{scope}[shift={(54,0)}]
		\draw[step=4cm,black,very thin] (0,0) grid (8,8);
		
		\begin{scope}[shift={(4,4)}]
			
			\begin{scope}[shift={(2,2)}]
				\fill[blue!50!black] (0,0) rectangle (1,1);
				\fill[blue!50!black] (1,1) rectangle (2,2);
				
				\draw[step=1cm,black,very thin] (0,0) grid (2,2);
			\end{scope}		
			
			\begin{scope}
				\fill[blue!50!black] (0,0) rectangle (1,1);
				\fill[blue!50!black] (1,1) rectangle (2,2);
				
				\draw[step=1cm,black,very thin] (0,0) grid (2,2);
			\end{scope}

			\draw[step=2cm,black,very thin] (0,0) grid (4,4);
		\end{scope}

		\begin{scope}[shift={(0,0)}]
			
			\begin{scope}[shift={(2,2)}]
				\fill[blue!50!black] (0,0) rectangle (1,1);
				\fill[blue!50!black] (1,1) rectangle (2,2);
				
				\draw[step=1cm,black,very thin] (0,0) grid (2,2);
			\end{scope}		
			
			\begin{scope}
				\fill[blue!50!black] (0,0) rectangle (1,1);
				\fill[blue!50!black] (1,1) rectangle (2,2);
				
				\draw[step=1cm,black,very thin] (0,0) grid (2,2);
			\end{scope}		
			
			\draw[step=2cm,black,very thin] (0,0) grid (4,4);
		\end{scope}

	\end{scope}

\end{scope}}}
		\end{aligned}\,.
	\end{align}	
	And so on.
	
	\section{Dimensional reduction in the adapted representations}\label{sec:DescRevisited}
	
	In this section the reduction procedure will be reviewed from the point of view of the adapted representations: we will recover the results of Section~\ref{sec:descent}, and underline the peculiarities with respect to the general case.
	
	It is important to point out that the algebraic and re\-pre\-sen\-ta\-tion-in\-de\-pen\-dent approach of Section~\ref{sec:descent}, and the constructive and re\-pre\-sen\-ta\-tion-de\-pen\-dent approach of this section are independent. Therefore, recovering the previous results also means providing \emph{alternative proofs} of such results.
	
	\subsection{Recovering the general results}
	Here we will show how the main nontrivial results of Section~\ref{sec:descent} can be recovered by working in the adapted representations. 
	
	As to the properties of the even-to-odd descent, we just have to set $d=2\ell$ ($\ell>1$), write the gamma matrices as in Eq.~\eqref{eq:complement},
	\begin{align}\label{eq:descready}
		\gamma_{(2\ell)}^{\mu'}
		=\left(\begin{array}{c|c}
			\gamma_{(2\ell-1)}^{\mu'}
			&\vphantom{\Bigl.\Bigr.}\\
			\hline
			\vphantom{\Bigl.\Bigr.}
			&	\gammac_{(2\ell-1)}^{\mu'}
		\end{array}\right)\,,
		&&
		\gamma_{(2\ell)}^{2\ell-1}
		=\left(\begin{array}{c|c}
			\vphantom{\Bigl.\Bigr.}
			&\iu\gamma_{(2\ell-1)}^{2\ell-2}\\
			\hline
			\iu\gammac_{(2\ell-1)}^{2\ell-2}
			&\vphantom{\Bigl.\Bigr.}
		\end{array}\right)
		\,,
	\end{align}
	and then everything follows more or less by direct inspection.
	
	To begin with, Eq.~\eqref{eq:descready} is an instance of the block-structure~\eqref{eq:BlockDiagReprGamma}, a built-in feature of the adapted representations, bringing about the decoupling by descent of the Dirac equation.
	We know from the start that the diagonal blocks of the $\smash{\cramped{\gamma_{(2\ell)}^{\mu'}}}$ determine two $(2\ell-1)$-dimensional inequivalent sets of gamma matrices, also satisfying Eq.~\eqref{eq:herm}. In particular, the inequivalence is now obtained by a visual proof (see Eqs.~\eqref{eq:oddInequi}--\eqref{eq:complement}). The block structure of $\smash{\cramped{\gamma_{(2\ell)}^{2\ell-1}}}$ is an instance of Eq.~\eqref{eq:blocksLast}, with real coefficients.
	
	The restricted covariance with respect to the proper and ortho\-chro\-nous $(2\ell-1)$-dimensional Lorentz transformations~\eqref{eq:lambda} follows by observing that such transformations are generated by $\frac{\iu}{2}\,\smash{\cramped{\gamma_{(2\ell)}^{\mu'}}}\smash{\cramped{\gamma_{(2\ell)}^{\nu'}}}$, $0\le\mu'<\nu'\le2\ell-2$, and that any such generator is block-diagonal (the remaining  generators in $2\ell$ dimensions involve $\gamma_{(2\ell)}^{2\ell-1}$, and  are all off-block-diagonal).
	
	The decoupling mechanism of Eqs.~\eqref{eq:EvenToOddDescent0}--\eqref{eq:EvenToOddDescent} is specialized in the following way. The Dirac equation written in the $2\ell$-dimensional adapted representation is
	\begin{align}\label{eq:adaptedDiracEven}
		\left(
		\begin{array}{c|c}
			\iu\gamma^{\mu'}_{(2\ell-1)}\partial_{\mu'}-m
			&
			-\gamma_{(2\ell-1)}^{2\ell-2}\,\partial_{2\ell-1}
			\vphantom{\biggl.\biggr.}\\
			\hline
			-\gammac_{(2\ell-1)}^{2\ell-2} \,\partial_{2\ell-1} \vphantom{\biggl.\biggr.}
			&
			\iu\gammac^{\mu'}_{(2\ell-1)}\partial_{\mu'}-m
		\end{array}\right)
		\Psi_{(2\ell)}
		=0\,.
	\end{align}
	Hereafter, a dimensional label will be attached to the wa\-ve\-func\-tions as well, to underline that they are expressed in some specific representation---the $2\ell$-dimensional adapted one.
	When the descent condition~\eqref{eq:DescentConditionEven} is applied, the equation splits up into two decoupled $(2\ell-1)$-dimensional Dirac equations, 
	\begin{align}\label{eq:adaptedred}
		\left(
		\begin{array}{c|c}
			\iu\gamma_{(2\ell-1)}^{\mu'}\partial_{\mu'}-m
			&\vphantom{\biggl.\biggr.}\\
			\hline
			\vphantom{\biggl.\biggr.}
			&\iu\gammac_{(2\ell-1)}^{\mu'}\partial_{\mu'}-m
		\end{array}\right)
		\Psi_{(2\ell)}
		=0\,,
	\end{align}		
	written in the complementary sets~\eqref{eq:constr_odd_1}--\eqref{eq:oddInequi}. After bringing out the $(2\ell-2)$-dimensional adapted matrices, the reduced equations are put in the form
	\begin{align}\label{eq:adaptedredbis}
		\qty(\iu\gamma^{\mu''}_{(2\ell-2)}\partial_{\mu''}
		\mp\gamma_{(2\ell-2)}^{\rm ch}\,\partial_{2\ell-2}-m)\Psi_{(2\ell-1)}=0\,.
	\end{align}
	The equations~\eqref{eq:adaptedredbis} are transformed into each other by inverting the last coordinate, $x^{2\ell-2}\to-x^{2\ell-2}$, a unitary transformation on the reduced wa\-ve\-func\-tion. This result generalizes what was already observed in~\cite{descent2} about the reduction of the ordinary Dirac equation.
	
	As to the odd-to-even descent, it is worth remarking that, when performing the descent on the reduced equations~\eqref{eq:adaptedredbis}, by imposing the independence of the last coordinate on the reduced wa\-ve\-func\-tions, that is $\partial_{2\ell-2}\Psi_{(2\ell-1)}=0$, the same equation is obtained, namely the Dirac equation in the $(2\ell-2)$-dimensional adapted representation,
	\begin{equation}\label{eq:afterdescent_oddtoeven_explicit}
		\left(\iu\gamma^{\mu''}_{(2\ell-2)}\partial_{\mu''}-m\right)\Psi_{(2\ell-2)}=0\,.
	\end{equation}
	
	\subsection{Visual descent in the adapted representations}
	
	By conceiving the dimensional reduction in terms of gamma matrices, the essential features of the even-to-odd and the odd-to-even descent are better illustrated side by side with a visual example. To this end, let us start with the even-to-odd descent, and consider the reduction of the 6-dimensional adapted set~\eqref{eq:6dimVerbo}--\eqref{eq:6dim}:
	\begin{align}
		\begin{aligned}
			\raisebox{-1.5pt}{\tikz{		\begin{scope}[scale=.27]

	\begin{scope}[shift={(0,0)}]
		
		\begin{scope}[shift={(0,4)}]
			
			\begin{scope}[shift={(0,2)}]
				\fill[red!50!black] (1,0) rectangle (2,1);
				\fill[blue!50!black] (0,1) rectangle (1,2);
				\draw[step=1cm,black,very thin] (0,0) grid (2,2);	
			\end{scope}		
			
			\begin{scope}[shift={(2,0)}]
				\fill[red!50!black] (1,0) rectangle (2,1);
				\fill[blue!50!black] (0,1) rectangle (1,2);
				\draw[step=1cm,black,very thin] (0,0) grid (2,2);	
			\end{scope}		
			\draw[step=2cm,black,very thin] (0,0) grid (4,4);
		\end{scope}	
		
		\begin{scope}[shift={(4,0)}]

			\begin{scope}[shift={(0,2)}]
				\fill[red!50!black] (1,0) rectangle (2,1);
				\fill[blue!50!black] (0,1) rectangle (1,2);
				\draw[step=1cm,black,very thin] (0,0) grid (2,2);	
			\end{scope}		
			
			\begin{scope}[shift={(2,0)}]
				\fill[red!50!black] (1,0) rectangle (2,1);
				\fill[blue!50!black] (0,1) rectangle (1,2);
				\draw[step=1cm,black,very thin] (0,0) grid (2,2);	
			\end{scope}		
			\draw[step=2cm,black,very thin] (0,0) grid (4,4);
		\end{scope}	
		
		\draw[step=4cm,black,very thin] (0,0) grid (8,8);
	\end{scope}

	\begin{scope}[shift={(9,0)}]
		
		\begin{scope}[shift={(0,4)}]
			
			\begin{scope}[shift={(0,2)}]
				\fill[red!50!black] (0,0) rectangle (1,1);
				\fill[blue!50!black] (1,1) rectangle (2,2);
				\draw[step=1cm,black,very thin] (0,0) grid (2,2);	
			\end{scope}		
			
			\begin{scope}[shift={(2,0)}]
				\fill[red!50!black] (0,0) rectangle (1,1);
				\fill[blue!50!black] (1,1) rectangle (2,2);
				\draw[step=1cm,black,very thin] (0,0) grid (2,2);	
			\end{scope}		
			\draw[step=2cm,black,very thin] (0,0) grid (4,4);
		\end{scope}	
		
		\begin{scope}[shift={(4,0)}]

			\begin{scope}[shift={(0,2)}]
				\fill[red!50!black] (0,0) rectangle (1,1);
				\fill[blue!50!black] (1,1) rectangle (2,2);
				\draw[step=1cm,black,very thin] (0,0) grid (2,2);	
			\end{scope}		
			
			\begin{scope}[shift={(2,0)}]
				\fill[red!50!black] (0,0) rectangle (1,1);
				\fill[blue!50!black] (1,1) rectangle (2,2);
				\draw[step=1cm,black,very thin] (0,0) grid (2,2);	
			\end{scope}		
			\draw[step=2cm,black,very thin] (0,0) grid (4,4);
		\end{scope}	
		
		\draw[step=4cm,black,very thin] (0,0) grid (8,8);
	\end{scope}

	\begin{scope}[shift={(18,0)}]
		
		\begin{scope}[shift={(0,4)}]
			
			\fill[orange!75!black] (0,2) rectangle (1,3);
			\fill[orange!75!black] (1,3) rectangle (2,4);

			\draw[step=1cm,black,very thin] (0,2) grid (2,4);
			
			\begin{scope}[shift={(2,-2)}]
				\fill[green!50!black] (0,2) rectangle (1,3);
				\fill[green!50!black] (1,3) rectangle (2,4);
				\draw[step=1cm,black,very thin] (0,2) grid (2,4);
			\end{scope}		
			
			\draw[step=2cm,black,very thin] (0,0) grid (4,4);		
			
		\end{scope}	
		
		\begin{scope}[shift={(4,0)}]

			\fill[orange!75!black] (0,2) rectangle (1,3);
			\fill[orange!75!black] (1,3) rectangle (2,4);

			\draw[step=1cm,black,very thin] (0,2) grid (2,4);
			
			\begin{scope}[shift={(2,-2)}]
				\fill[green!50!black] (0,2) rectangle (1,3);
				\fill[green!50!black] (1,3) rectangle (2,4);
				\draw[step=1cm,black,very thin] (0,2) grid (2,4);
			\end{scope}		
			
			\draw[step=2cm,black,very thin] (0,0) grid (4,4);		
		\end{scope}	
		
		\draw[step=4cm,black,very thin] (0,0) grid (8,8);
	\end{scope}

	
	\begin{scope}[shift={(27,0)}]
		
		\begin{scope}[shift={(0,4)}]

			\begin{scope}[shift={(2,2)}]
				\fill[blue!50!black] (0,0) rectangle (1,1);
				\fill[blue!50!black] (1,1) rectangle (2,2);
				
				\draw[step=1cm,black,very thin] (0,0) grid (2,2);
			\end{scope}		
			
			\begin{scope}
				\fill[red!50!black] (0,0) rectangle (1,1);
				\fill[red!50!black] (1,1) rectangle (2,2);
				
				\draw[step=1cm,black,very thin] (0,0) grid (2,2);
			\end{scope}

			\draw[step=2cm,black,very thin] (0,0) grid (4,4);
			
		\end{scope}	
		
		\begin{scope}[shift={(4,0)}]

			\begin{scope}[shift={(2,2)}]
				\fill[blue!50!black] (0,0) rectangle (1,1);
				\fill[blue!50!black] (1,1) rectangle (2,2);
				
				\draw[step=1cm,black,very thin] (0,0) grid (2,2);
			\end{scope}		
			
			\begin{scope}
				\fill[red!50!black] (0,0) rectangle (1,1);
				\fill[red!50!black] (1,1) rectangle (2,2);
				
				\draw[step=1cm,black,very thin] (0,0) grid (2,2);
			\end{scope}

			\draw[step=2cm,black,very thin] (0,0) grid (4,4);
		\end{scope}	
		
		\draw[step=4cm,black,very thin] (0,0) grid (8,8);
	\end{scope}

	\begin{scope}[shift={(36,0)}]

		\begin{scope}[shift={(0,4)}]

			\fill[orange!75!black] (0,0) rectangle (1,1);
			\fill[orange!75!black] (1,1) rectangle (2,2);

			\draw[step=1cm,black,very thin] (0,0) grid (2,2);
			
			\begin{scope}[shift={(2,2)}]
				\fill[orange!75!black] (0,0) rectangle (1,1);
				\fill[orange!75!black] (1,1) rectangle (2,2);
				
				\draw[step=1cm,black,very thin] (0,0) grid (2,2);
				
			\end{scope}

			\draw[step=2cm,black,very thin] (0,0) grid (4,4);
		\end{scope}	
		
		\begin{scope}[shift={(4,0)}]

			\fill[green!50!black] (0,0) rectangle (1,1);
			\fill[green!50!black] (1,1) rectangle (2,2);

			\draw[step=1cm,black,very thin] (0,0) grid (2,2);
			
			\begin{scope}[shift={(2,2)}]
				\fill[green!50!black] (0,0) rectangle (1,1);
				\fill[green!50!black] (1,1) rectangle (2,2);
				
				\draw[step=1cm,black,very thin] (0,0) grid (2,2);
				
			\end{scope}		
			
			\draw[step=2cm,black,very thin] (0,0) grid (4,4);
			
		\end{scope}			
		
		\draw[step=4cm,black,very thin] (0,0) grid (8,8);
	\end{scope}

	
	\begin{scope}[shift={(45,0)}]
		\draw[step=4cm,black,very thin] (0,0) grid (8,8);
		
		\begin{scope}[shift={(4,4)}]
			
			\begin{scope}[shift={(2,2)}]
				\fill[blue!50!black] (0,0) rectangle (1,1);
				\fill[blue!50!black] (1,1) rectangle (2,2);
				
				\draw[step=1cm,black,very thin] (0,0) grid (2,2);
			\end{scope}		
			
			\begin{scope}
				\fill[blue!50!black] (0,0) rectangle (1,1);
				\fill[blue!50!black] (1,1) rectangle (2,2);
				
				\draw[step=1cm,black,very thin] (0,0) grid (2,2);
			\end{scope}

			\draw[step=2cm,black,very thin] (0,0) grid (4,4);
		\end{scope}

		\begin{scope}[shift={(0,0)}]
			
			\begin{scope}[shift={(2,2)}]
				\fill[red!50!black] (0,0) rectangle (1,1);
				\fill[red!50!black] (1,1) rectangle (2,2);
				
				\draw[step=1cm,black,very thin] (0,0) grid (2,2);
			\end{scope}		
			
			\begin{scope}
				\fill[red!50!black] (0,0) rectangle (1,1);
				\fill[red!50!black] (1,1) rectangle (2,2);
				
				\draw[step=1cm,black,very thin] (0,0) grid (2,2);
			\end{scope}		
			
			\draw[step=2cm,black,very thin] (0,0) grid (4,4);
		\end{scope}

	\end{scope}

\end{scope}}}
		\end{aligned}\,.
	\end{align}
	The even-to-odd descent in the adapted representations simply consists in dropping the last matrix, and grouping together the upper-left and the lower-right blocks of the remaining block-diagonal matrices: we get two inequivalent odd sets, differing by the sign of the last matrix, and the one coming from the upper-left blocks is the underlying (odd) adapted set.
	
	In our specific case, the set of the upper-left $4\times4$ blocks,
	\begin{align}\label{eq:5red}
		\begin{aligned}
			\raisebox{-1.5pt}{\tikz{\begin{scope}[scale=.36]

	
	\begin{scope}[shift={(0,2)}]
		\fill[red!50!black] (1,0) rectangle (2,1);
		\fill[blue!50!black] (0,1) rectangle (1,2);
		
		\draw[step=1cm,black,very thin] (0,0) grid (2,2);
	\end{scope}		
	
	\begin{scope}[shift={(2,0)}]
		\fill[red!50!black] (1,0) rectangle (2,1);
		\fill[blue!50!black] (0,1) rectangle (1,2);
		
		\draw[step=1cm,black,very thin] (0,0) grid (2,2);
	\end{scope}		
	
	\draw[step=2cm,black,very thin] (0,0) grid (4,4);

	\begin{scope}[shift={(5,0)}]

		\fill[red!50!black] (0,2) rectangle (1,3);
		\fill[blue!50!black] (1,3) rectangle (2,4);
		
		\draw[step=1cm,black,very thin] (0,2) grid (2,4);
		
		\begin{scope}[shift={(2,-2)}]
			\fill[red!50!black] (0,2) rectangle (1,3);
			\fill[blue!50!black] (1,3) rectangle (2,4);
			
			\draw[step=1cm,black,very thin] (0,2) grid (2,4);
		\end{scope}		
		
		\draw[step=2cm,black,very thin] (0,0) grid (4,4);

	\end{scope}

	\begin{scope}[shift={(10,0)}]
		
		\fill[orange!75!black] (0,2) rectangle (1,3);
		\fill[orange!75!black] (1,3) rectangle (2,4);

		\draw[step=1cm,black,very thin] (0,2) grid (2,4);
		
		\begin{scope}[shift={(2,-2)}]
			\fill[green!50!black] (0,2) rectangle (1,3);
			\fill[green!50!black] (1,3) rectangle (2,4);
			\draw[step=1cm,black,very thin] (0,2) grid (2,4);
		\end{scope}		
		
		\draw[step=2cm,black,very thin] (0,0) grid (4,4);		
		
	\end{scope}

	\begin{scope}[shift={(15,0)}]
		
		\begin{scope}[shift={(2,2)}]
			\fill[blue!50!black] (0,0) rectangle (1,1);
			\fill[blue!50!black] (1,1) rectangle (2,2);
			
			\draw[step=1cm,black,very thin] (0,0) grid (2,2);
		\end{scope}		
		
		\begin{scope}
			\fill[red!50!black] (0,0) rectangle (1,1);
			\fill[red!50!black] (1,1) rectangle (2,2);
			
			\draw[step=1cm,black,very thin] (0,0) grid (2,2);
		\end{scope}

		\draw[step=2cm,black,very thin] (0,0) grid (4,4);

	\end{scope}

	\begin{scope}[shift={(20,0)}]

		\begin{scope}[shift={(2,2)}]
			\fill[orange!75!black] (0,0) rectangle (1,1);
			\fill[orange!75!black] (1,1) rectangle (2,2);
			
			\draw[step=1cm,black,very thin] (0,0) grid (2,2);
		\end{scope}		
		
		\begin{scope}
			\fill[orange!75!black] (0,0) rectangle (1,1);
			\fill[orange!75!black] (1,1) rectangle (2,2);
			
			\draw[step=1cm,black,very thin] (0,0) grid (2,2);
		\end{scope}

		\draw[step=2cm,black,very thin] (0,0) grid (4,4);
		
	\end{scope}

\end{scope}}}
		\end{aligned}\,,
	\end{align}
	is indeed the 5-dimensional adapted set~\eqref{eq:5dimVerbo}--\eqref{eq:5dim}, characterized by $\smash{\cramped{\gamma_{(5)}^{4}}}=-\iu\smash{\cramped{\gamma_{(4)}^{\rm ch}}}$, and corresponding to the choice we made when defining the odd adapted sets. The lower-right set is of course complementary to the previous one, 
	\begin{align}\label{eq:anti5red}
		\begin{aligned}
			\raisebox{-1.5pt}{\tikz{\begin{scope}[scale=.36]

	
	\begin{scope}[shift={(0,2)}]
		\fill[red!50!black] (1,0) rectangle (2,1);
		\fill[blue!50!black] (0,1) rectangle (1,2);
		
		\draw[step=1cm,black,very thin] (0,0) grid (2,2);
	\end{scope}		
	
	\begin{scope}[shift={(2,0)}]
		\fill[red!50!black] (1,0) rectangle (2,1);
		\fill[blue!50!black] (0,1) rectangle (1,2);
		
		\draw[step=1cm,black,very thin] (0,0) grid (2,2);
	\end{scope}		
	
	\draw[step=2cm,black,very thin] (0,0) grid (4,4);

	\begin{scope}[shift={(5,0)}]

		\fill[red!50!black] (0,2) rectangle (1,3);
		\fill[blue!50!black] (1,3) rectangle (2,4);
		
		\draw[step=1cm,black,very thin] (0,2) grid (2,4);
		
		\begin{scope}[shift={(2,-2)}]
			\fill[red!50!black] (0,2) rectangle (1,3);
			\fill[blue!50!black] (1,3) rectangle (2,4);
			
			\draw[step=1cm,black,very thin] (0,2) grid (2,4);
		\end{scope}		
		
		\draw[step=2cm,black,very thin] (0,0) grid (4,4);

	\end{scope}

	\begin{scope}[shift={(10,0)}]
		
		\fill[orange!75!black] (0,2) rectangle (1,3);
		\fill[orange!75!black] (1,3) rectangle (2,4);

		\draw[step=1cm,black,very thin] (0,2) grid (2,4);
		
		\begin{scope}[shift={(2,-2)}]
			\fill[green!50!black] (0,2) rectangle (1,3);
			\fill[green!50!black] (1,3) rectangle (2,4);
			\draw[step=1cm,black,very thin] (0,2) grid (2,4);
		\end{scope}		
		
		\draw[step=2cm,black,very thin] (0,0) grid (4,4);		
		
	\end{scope}

	\begin{scope}[shift={(15,0)}]
		
		\begin{scope}[shift={(2,2)}]
			\fill[blue!50!black] (0,0) rectangle (1,1);
			\fill[blue!50!black] (1,1) rectangle (2,2);
			
			\draw[step=1cm,black,very thin] (0,0) grid (2,2);
		\end{scope}		
		
		\begin{scope}
			\fill[red!50!black] (0,0) rectangle (1,1);
			\fill[red!50!black] (1,1) rectangle (2,2);
			
			\draw[step=1cm,black,very thin] (0,0) grid (2,2);
		\end{scope}

		\draw[step=2cm,black,very thin] (0,0) grid (4,4);

	\end{scope}

	\begin{scope}[shift={(20,0)}]

		\begin{scope}[shift={(2,2)}]
			\fill[green!50!black] (0,0) rectangle (1,1);
			\fill[green!50!black] (1,1) rectangle (2,2);
			
			\draw[step=1cm,black,very thin] (0,0) grid (2,2);
		\end{scope}		
		
		\begin{scope}
			\fill[green!50!black] (0,0) rectangle (1,1);
			\fill[green!50!black] (1,1) rectangle (2,2);
			
			\draw[step=1cm,black,very thin] (0,0) grid (2,2);
		\end{scope}

		\draw[step=2cm,black,very thin] (0,0) grid (4,4);
		
	\end{scope}

\end{scope}}}
		\end{aligned}\,,
	\end{align}
	from which it is obtained by changing the sign to the last matrix. 
	
	The odd-to-even descent always consists in dropping the last gamma matrix. In a adapted representation, the end result is always the underlying (even) adapted set. In our case, the sets~\eqref{eq:5red} and~\eqref{eq:anti5red} both reduce to the 4-dimensional  adapted set~\eqref{eq:4dimVerbo}--\eqref{eq:4dim}:
	\begin{align}
		\begin{aligned}
			\raisebox{-1.5pt}{\tikz{		\begin{scope}[scale=.36]
	
	
	\begin{scope}[shift={(0,2)}]
		\fill[red!50!black] (1,0) rectangle (2,1);
		\fill[blue!50!black] (0,1) rectangle (1,2);
		
		\draw[step=1cm,black,very thin] (0,0) grid (2,2);
	\end{scope}		
	
	\begin{scope}[shift={(2,0)}]
		\fill[red!50!black] (1,0) rectangle (2,1);
		\fill[blue!50!black] (0,1) rectangle (1,2);
		
		\draw[step=1cm,black,very thin] (0,0) grid (2,2);
	\end{scope}		
	
	\draw[step=2cm,black,very thin] (0,0) grid (4,4);

	\begin{scope}[shift={(5,0)}]

		\fill[red!50!black] (0,2) rectangle (1,3);
		\fill[blue!50!black] (1,3) rectangle (2,4);
		
		\draw[step=1cm,black,very thin] (0,2) grid (2,4);
		
		\begin{scope}[shift={(2,-2)}]
			\fill[red!50!black] (0,2) rectangle (1,3);
			\fill[blue!50!black] (1,3) rectangle (2,4);
			
			\draw[step=1cm,black,very thin] (0,2) grid (2,4);
		\end{scope}		
		
		\draw[step=2cm,black,very thin] (0,0) grid (4,4);

	\end{scope}

	\begin{scope}[shift={(10,0)}]
		
		\fill[orange!75!black] (0,2) rectangle (1,3);
		\fill[orange!75!black] (1,3) rectangle (2,4);

		\draw[step=1cm,black,very thin] (0,2) grid (2,4);
		
		\begin{scope}[shift={(2,-2)}]
			\fill[green!50!black] (0,2) rectangle (1,3);
			\fill[green!50!black] (1,3) rectangle (2,4);
			\draw[step=1cm,black,very thin] (0,2) grid (2,4);
		\end{scope}		
		
		\draw[step=2cm,black,very thin] (0,0) grid (4,4);		
		
	\end{scope}

	\begin{scope}[shift={(15,0)}]
		
		\begin{scope}[shift={(2,2)}]
			\fill[blue!50!black] (0,0) rectangle (1,1);
			\fill[blue!50!black] (1,1) rectangle (2,2);
			
			\draw[step=1cm,black,very thin] (0,0) grid (2,2);
		\end{scope}		
		
		\begin{scope}
			\fill[red!50!black] (0,0) rectangle (1,1);
			\fill[red!50!black] (1,1) rectangle (2,2);
			
			\draw[step=1cm,black,very thin] (0,0) grid (2,2);
		\end{scope}

		\draw[step=2cm,black,very thin] (0,0) grid (4,4);

	\end{scope}

\end{scope}}}
		\end{aligned}\,.
	\end{align}
	Then, it is easy to see that a new descent brings about the reduced sets
	\begin{align}\label{eq:3red}
		&	\begin{aligned}
			\raisebox{-1.5pt}{\tikz{	\begin{scope}[scale=.36]

	\fill[white] (0,2) rectangle (2,0);
	\fill[red!50!black] (1,0) rectangle (2,1);
	\fill[blue!50!black] (0,1) rectangle (1,2);			
	
	\draw[step=1cm,black,very thin] (0,0) grid (2,2);	
	
	\begin{scope}[shift={(3,0)}]
		
		\fill[white] (0,2) rectangle (2,0);
		\fill[red!50!black] (0,0) rectangle (1,1);
		\fill[blue!50!black] (1,1) rectangle (2,2);
		
		\draw[step=1cm,black,very thin] (0,0) grid (2,2);
	\end{scope}

	\begin{scope}[shift={(6,0)}]
		
		\fill[orange!75!black] (0,0) rectangle (1,1);
		\fill[orange!75!black] (1,1) rectangle (2,2);
		
		\draw[step=1cm,black,very thin] (0,0) grid (2,2);
	\end{scope}

\end{scope}}}
		\end{aligned}\,,\\
		&	\begin{aligned}
			\raisebox{-1.5pt}{\tikz{	\begin{scope}[scale=.36]

	\fill[white] (0,2) rectangle (2,0);
	\fill[red!50!black] (1,0) rectangle (2,1);
	\fill[blue!50!black] (0,1) rectangle (1,2);			
	
	\draw[step=1cm,black,very thin] (0,0) grid (2,2);	
	
	\begin{scope}[shift={(3,0)}]
		
		\fill[white] (0,2) rectangle (2,0);
		\fill[red!50!black] (0,0) rectangle (1,1);
		\fill[blue!50!black] (1,1) rectangle (2,2);
		
		\draw[step=1cm,black,very thin] (0,0) grid (2,2);
	\end{scope}

	\begin{scope}[shift={(6,0)}]
		
		\fill[green!50!black] (0,0) rectangle (1,1);
		\fill[green!50!black] (1,1) rectangle (2,2);
		
		\draw[step=1cm,black,very thin] (0,0) grid (2,2);
	\end{scope}

\end{scope}}}
		\end{aligned}\,,\label{eq:anti3red}
	\end{align}
	the set~\eqref{eq:3red} being the 3-dimensional adapted set~\eqref{eq:3dimVerbo}--\eqref{eq:3dim}, and, by a further descent, both are finally reduced to the 2-dimensional adapted set~\eqref{eq:2dimVerbo}--\eqref{eq:2dim}, 
	\begin{align}
		\begin{aligned}
			\raisebox{-1.5pt}{\tikz{\begin{scope}[scale=.36]

	\fill[white] (0,2) rectangle (2,0);
	\fill[red!50!black] (1,0) rectangle (2,1);
	\fill[blue!50!black] (0,1) rectangle (1,2);			
	
	\draw[step=1cm,black,very thin] (0,0) grid (2,2);	
	
	\begin{scope}[shift={(3,0)}]
		
		\fill[white] (0,2) rectangle (2,0);
		\fill[red!50!black] (0,0) rectangle (1,1);
		\fill[blue!50!black] (1,1) rectangle (2,2);
		
		\draw[step=1cm,black,very thin] (0,0) grid (2,2);
	\end{scope}

\end{scope}}}
		\end{aligned}\,.
	\end{align}
	
	In general, the iterated even-to-even descent can be characterized by the tree diagram below, where $d=2\ell$ ($\ell>1$) is the starting dimension, and the adapted sequence is in red:
	\begin{align}\label{eq:albero}
		\begin{aligned}
				\begin{tikzpicture}[shorten >=1pt,node distance=80pt,auto]
	\node[state,style={fill=red!15,draw=red!50!black,thick,circular drop shadow}] (E) 
	{$\mathmakebox[\widthof{$\gammac^{\mu'}_{(2\ell-1)}$}]{\gamma^{\mu'}_{(2\ell-1)}}$};
	\node[state,style={fill=red!20,draw=red!50!black,thick,circular drop shadow}] (N) [above right of=E] {$\mathmakebox[\widthof{$\gammac^{\mu'}_{(2\ell-1)}$}]{\gamma^{\mu}_{(2\ell)}}
		\vphantom{\gammac^{\mu'}_{(2\ell-1)}}$};
	\node[state,style={fill=red!10,draw=red!50!black,thick,circular drop shadow}] (S) [below right of=E] {$\mathmakebox[\widthof{$\gammac^{\mu'}_{(2\ell-1)}$}]{\gamma^{\mu''}_{(2\ell-2)}}$};
	\node[state,style={fill=black!10,circular drop shadow}] (W) [below right of=N] {$\mathmakebox[\widthof{$\gammac^{\mu'}_{(2\ell-1)}$}]{\gammac^{\mu'}_{(2\ell-1)}}$};
	\path[->,style={draw=red!50!black}, thick] (N) edge  (E) 
	(E) edge (S);
		\path[->] (N) edge     (W)
		(W) edge (S);
\end{tikzpicture}
		\end{aligned}\;.
	\end{align}
	The considerations of Subsections~\ref{subsec:evenConstruction} and~\ref{subsec:oddConstruction} show that the $2\ell$-dimensional adapted set has the $(2\ell-2)$-dimensional adapted set as its only grandchild: the two branches stemming out of each even set always come together in correspondence of the underlying even set.
	
	\section{Concluding remarks}\label{sec:conclusions}
	
	We have shown that the whole hierarchy of Dirac equations in arbitrary spatial dimensions can be derived through the prescription of dimensional reduction. The pattern is the following: by performing the dimensional reduction on the Dirac theory in an odd spacetime, one obtains the Dirac theory in the lower (even)-dimensional spacetime. By contrast, starting from an even spacetime, one obtains a system of two decoupled inequivalent Dirac theories, each one obtained as the restriction of the original Dirac theory to the eigenspace of a suitable superselection charge. We introduced a family of representations of the gamma matrices in which the latter is diagonal: by adopting such representations, the decoupled Dirac theories obtained by performing dimensional reduction on the Dirac equation on an even-dimensional spacetime are mapped into each other via reflection of one spatial coordinate. This generalizes the results already obtained in~\cite{descent2} for the Dirac equations in $(3+1)$- and $(2+1)$-dimensional spacetimes. A further descent along the last spatial coordinate finally yields two equivalent theories.
	
	A few comments are in order about the different behavior when one descends from odd to even dimensions and \emph{vice versa}. In odd spatial dimensions, the inversion of all coordinates swaps the two different orientations, while in even dimensions it leaves the orientation unchanged. Therefore, if we start from odd dimensions and remove one coordinate, what is left in even dimensions has the same orientation, regardless of the orientation of the higher-dimensional space. By contrast, if we start from even dimensions  and remove one coordinate,  what is left, in odd dimensions, can have different orientations. 
	
	\section*{Acknowledgments}
	
	This research was funded by MIUR via PRIN 2017 (Pro\-get\-to di Ri\-cer\-ca di In\-te\-res\-se Na\-zio\-na\-le), project QUSHIP (2017SRNBRK), by the Italian National Group of Mathematical Physics (GNFM-INdAM), by I\-sti\-tu\-to Na\-zio\-na\-le di Fi\-si\-ca Nu\-cle\-are (INFN) through the project ``QUANTUM'', and by Re\-gio\-ne Pu\-glia and QuantERA ERA-NET Cofund in Quantum Technologies (Grant No. 731473), project QuantHEP.


\begin{thebibliography}{99}
		
		\small 
		
		\bibitem{wigner} E. P. Wigner, ``The unreasonable effectiveness of mathematics in the natural sciences. Richard Courant lecture in mathematical sciences delivered at New York University, May 11, 1959''. Communications on Pure and Applied Mathematics \textbf{13}, 1--14 (1960).
		
		\bibitem{dirac} P. A. M. Dirac, \textit{Principles of Quantum Mechanics}. International Series of Monographs on Physics (4th ed.), Oxford University Press, p. 255, 1958.
		
		\bibitem{Weinberg} S. Weinberg, \textit{The Quantum Theory of Fields}, Cambridge: Cambridge University Press, 1995. 
		
		\bibitem{ehrenfest} P. Ehrenfest, ``In what way does it become manifest in the fundamental laws of physics that space has three dimensions?''. Proceedings of the Royal Academy of Sciences at Amsterdam \textbf{20}, 200--209 (1918).
		
		\bibitem{history} P. A. M. Dirac, ``Recollections of an exciting era'', in \emph{History of Twentieth-Century Physics}, Proceedings of the International School of Physics ``Enrico Fermi'', Course 57, edited by C. Weiner, Academic Press, New York and London, 1977.
		
		\bibitem{dirac28} P. A. M. Dirac, ``The Quantum Theory of the Electron''. Proceedings of the Royal Society A: Mathematical, Physical and Engineering Sciences \textbf{117}, 610--624 (1928).		
		
		\bibitem{deWitSmith} B.~de~Wit, J.~Smith, \textit{Field Theory in Particle Physics, Volume 1}, North Holland, 1986.			
		
		\bibitem{Cornwell} J.~F.~Cornwell, \textit{Group Theory in Physics, Volume 3}, Academic Press, 1989.	
		
		\bibitem{Simon} B.~Simon, \textit{Representations of Finite and Compact Groups}, American Mathematical Soc., 1996.
		
		\bibitem{kawada} Y. Kawada, N. Iwahori. ``On the structure and representations of Clifford algebras''. Journal of the Mathematical Society of Japan,  \textbf{2} (1-2), 34--43 (1950).
		
		\bibitem{pais} A.~Pais, ``On Spinors in $n$ Dimensions''. J. Math. Phys. \textbf{3}, 1135 (1962).
		
		\bibitem{hadamard} J. Hadamard, \textit{Lectures on Cauchy's problem in linear partial differential equations}. Yale University Press, New Haven, 1923.
		
		\bibitem{ehrenulen} P. Ehrenfest, and G. E. Uhlenbeck, ``On the Connection of Different Methods of Solution of the Wave Equation in Multi-dimensional Spaces''. Proceedings of the Royal Academy of Sciences at Amsterdam \textbf{29}, 1280 (1926).	
		
		\bibitem{balasz} N. L. Balazs, ``Wave Propagation in Even and Odd Dimensional Spaces''. Proceedings of the Physical Society Section A \textbf{68}, 521 (1955).		
		
		\bibitem{descent} R. Maggi, E. Ercolessi, P. Facchi, G. Marmo, S. Pascazio, and F. V. Pepe, ``Dimensional reduction of electromagnetism''. Journal of Mathematical Physics \textbf{63}, 022902 (2022).
		
		\bibitem{descent2} G. Angelone, R. Maggi, E. Ercolessi, P. Facchi, D. Lonigro, G. Marmo, S. Pascazio, and F. V. Pepe, ``Dimensional reduction of the Dirac theory''. arXiv:2211.08581 \texttt{[quant-ph]} (2022).
		
		\bibitem{BrauerWeyl} R.~Brauer, and H.~Weyl, ``Spinors in $n$ dimensions''. Am. J. Math. \textbf{57}, 425--449 (1935). 
				
		\bibitem{Lounesto} P.~Lounesto, \textit{Clifford Algebras and Spinors}, 2nd ed., London Mathematical Society Lecture Note Series, Cambridge: Cambridge University Press, 2001. 
		
		\bibitem{SilvaRocha} J. M. Hoff da Silva, R. da Rocha, ``Unfolding physics from the algebraic classification of spinor fields''. Phys. Lett. B \textbf{718}, 1519--1523 (2013). 
		
		\bibitem{Lee1} H.~C.~Lee, ``On Clifford's algebra''. London Math. Soc., Jour. \textbf{20}, 27--32 (1945).
		
		\bibitem{Lee} H.~C.~Lee, ``On Clifford Algebras and Their Representations''. Ann. Math. \textbf{49}, 760--773 (1948). 
		
		\bibitem{Go55} R. H. Good, ``Properties of the Dirac Matrices''. Rev. Mod. Phys. \textbf{27}, 187 (1955).
		
		\bibitem{park} J.~Park, ``Lecture note on Clifford algebra''. J. Korean Phys. Soc. \textbf{81}, 1--17 (2022). 
		
		\bibitem{Littlewood} D.~E.~Littlewood, ``Note on the Anticommuting Matrices of Eddington''. J. Lond. Math. Soc. \textbf{9}, 41 (1934)
		
		\bibitem{thaller} B. Thaller, \textit{The Dirac Equation}. Springer-Verlag Berlin Heidelberg, 1992.	
		
	\end{thebibliography}
\end{document}